\documentclass{aa}  

\usepackage{txfonts}
\usepackage[pdfpagelabels=false,colorlinks=true,linkcolor=blue,citecolor=blue,filecolor=blue,urlcolor=blue]{hyperref}
\usepackage[utf8]{inputenc}
\usepackage{amsmath}
\usepackage{graphicx}
\usepackage{float}
\usepackage{placeins}
\usepackage{macros_vdb}
\usepackage{mathrsfs}
\usepackage{color}
\usepackage{xcolor} 
\usepackage{soul}

\defcitealias{krumholz10}{KB10}
\defcitealias{krumholz18}{K18}
\defcitealias{dekel20}{D20}
\defcitealias{dekel09}{DSC09}
\defcitealias{dekel13}{D13}

\newcommand{\se}[1]{\S\ref{sec:#1}}

\begin{document} 

\title{Radial Transport in High Redshift Disk Galaxies Dominated by Inflowing Streams}

\authorrunning{Dutta Chowdhury et al.} \author{Dhruba Dutta Chowdhury \inst{1}\thanks{E-mail: dhruba.duttachowdhury@mail.huji.ac.il}, Avishai Dekel \inst{1, 2}, Nir Mandelker \inst{1}, Omri Ginzburg \inst{1}, and Reinhard Genzel \inst{3,4}}

\institute{Racah Institute of Physics, The Hebrew University of Jerusalem, Jerusalem 91904, Israel \and 
Santa Cruz Institute for Particle Physics, University of California, Santa Cruz, CA 95064, USA \and 
Max-Planck-Institut f\"{u}r Extraterrestrische Physik (MPE), Giessenbachstraße 1, D-85748 Garching, Germany \and
Departments of Physics and Astronomy, University of California, Berkeley, CA 94720, USA}

\date{Submitted: September 3, 2024; Revised: Aug 31, 2025; Accepted: Sep 29, 2025}

\abstract
{}
{We study the radial transport of cold gas within simulated disk galaxies at cosmic noon, aiming at distinguishing between disk instability and accretion along cold streams from the cosmic web as its driving mechanism.
}  
{Disks are selected based on kinematics and flattening from the VELA zoom-in hydro-cosmological simulations. The radial velocity fields in the disks are mapped, their averages are computed as a function of radius and over the whole disk, and the radial mass flux in each disk as a function of radius is obtained. The transport directly associated with fresh incoming streams is identified by selecting cold gas cells that are either on incoming streamlines or have low metallicity.
}
{The radial velocity fields in VELA disks are found to be highly non-axisymmetric, showing both inflows and outflows. However, in most cases, the average radial velocities, both as a function of radius and over the whole disk, are directed inwards, with the disk-averaged radial velocities typically amounting to a few percent of the disk-averaged rotational velocities. This is significantly lower than the expectations from various models that analytically predict the inward mass transport as driven by torques associated with disk instability. Under certain simplifying assumptions, the latter typically predict average inflows of more than $10\%$ of the rotational velocities. Analyzing the radial motions of streams and off-stream material, we find that the radial inflow in VELA disks is dominated by the stream inflows themselves, especially in the outer disks.
}
{The high inward radial velocities inferred in observed disks at cosmic noon, at the level of $\sim \! 20\%$ of the rotational velocities, may reflect inflowing streams from the cosmic web rather than being generated by disk instability.
} 


\maketitle

\section{Introduction}

Extended, self-gravitating, star-forming, highly turbulent, rotating disks dominate galaxy formation at cosmic noon, $z \sim 3-1$ \citep[e.g.,][]{erb04, forsterschreiber06, genzel06, elmegreen07, kassin07, genzel08, stark08, foresterschreiber09, kassin12, swinbank12, wisnioski15, stott16, simons17, swinbank17, ubler19, wisnioski19}, representing the peak of the star formation rate density in the Universe \citep[][]{madau14}.

Disks at cosmic noon undergo various instabilities, collectively known as violent disk instabilities \citep[VDI, e.g.,][]{dekel09}. \citet{toomre64} showed that a self-gravitating, rotating disk becomes gravitationally unstable when the Toomre-Q parameter, $Q \propto \sigma \kappa / \Sigma$, which represents the balance between self-gravity on the one hand (represented here by the gas surface density, $\Sigma$) and turbulent pressure and centrifugal support on the other hand (represented by the gas velocity dispersion, $\sigma$, and the epicyclic frequency, $\kappa$, respectively), becomes less than a critical value, $Q_{\rmc}$ of order unity. During the VDI phase, a disk can maintain a self-regulated, marginally unstable steady state with $Q \sim Q_{\rmc}$ \citep[e.g.,][]{dekel09, krumholz10, cacciato12, genel12a, forbes12, forbes14, krumholz18, ginzburg22}. If the disk becomes gravitationally unstable, the instabilities break axisymmetry and induce torques that lead to the transport of angular momentum radially outwards, which is compensated by the inward migration of mass until marginal instability is restored. As mass moves radially inwards down the potential well, the gravitational energy gained can be converted to turbulent energy, thereby maintaining the high levels of turbulence within the disk \citep[][]{wada02}. In support of the above self-regulation mechanism, observed disks at cosmic noon also tend to have $Q$ values of order unity \citep[e.g.,][]{genzel11,romeo13,genzel14,obreschkow15}.

VDI-driven radial mass transport can manifest in different ways. In \citet{dekel09}, clumps forming in a marginally unstable disk migrate inwards due to torques from encounters with the off-clump material and other clumps \citep[see also][]{cacciato12}. Multiple, giant, star-forming clumps have been observed in cosmic-noon disks \citep[e.g.,][]{elmegreen05, genzel11, shibuya16, huertas-company20, ginzburg21} and are also found to emerge from the gravitational fragmentation of gas-rich disks in isolated as well as cosmological simulations \citep[e.g.,][]{noguchi99, agertz09, genel12b, bournaud14, mandelker14, oklopvcic17, mandelker17}. In addition, inward clump migration can be fueled by dynamical friction on the clumps from the off-clump material, as shown in \citet[][]{dekel13}. 

Another mechanism is turbulent viscosity. According to \citet[][]{krumholz10}, turbulence can be modeled as an effective viscosity, which induces torques in a marginally unstable disk. The gravitational energy gained from the resulting inward mass transport balances the energy lost due to dissipation of turbulence such that an energy equilibrium is maintained \citep[see also][]{goldbaum15}. \citet[][]{krumholz18} extend the \citet[][]{krumholz10} model, considering turbulence driven by both disk instability and supernova feedback, where energy lost due to dissipation of turbulence is balanced by both energy gained from supernova feedback and VDI-driven radial mass transport in a marginally unstable disk. They find that while supernova-feedback-driven turbulence is sufficient to explain the observed velocity dispersions in low mass galaxies at low redshifts, disk-instability-driven turbulence is required at high redshifts \citep[see also][]{forbes12, forbes14, goldbaum16}. The \citet[][]{krumholz18} model is extended by \citet{ginzburg22}, who include an additional energy input from external accretion and show that while supernova feedback is the main driver of turbulence in low mass disks (halo masses $< 10^{11.5} M_{\odot}$ at $z \sim 2$), the primary driver of turbulence in high mass disks is either external accretion or VDI-based radial mass transport, depending on the efficiency by which stream kinetic energy is converted to turbulent energy, which is treated as a free parameter \citep[see also][]{elmegreen10, klessen10, genel12a, forbes23}. 

Apart from clump migration and viscous disk transport, VDI-driven radial mass transport in cosmic-noon disks can also be caused by the inward migration of long-lived rings due to torques from the rest of the disk. As shown in \citet{dekel20}, such rings can form via gravitational collapse in galaxies exhibiting a massive bulge.

For nearby galaxies, radial motions of a few km/s have been consistently inferred using H$\alpha$/CO observations \citep{wong04,sellwood10,schmidt16,speights19}. Similar studies in the distant Universe are difficult because of high resolution and signal-to-noise requirements. However, in a recent pioneering study, observational estimates of radial inflows in disks at cosmic noon were obtained by \citet{genzel23}. Using high-resolution H$\alpha$/CO imaging spectroscopy of nine moderately large, well-resolved, rotating disks between $z \sim 2.5-1$, \citet{genzel23} infer radially inward motions at the level of $50-100\ {\rm km/s}$ or at least $20 \%$ of the rotational velocities, which seem to be in the ballpark of the model predictions based on VDI. However, there are large observational uncertainties in the analysis. Therefore, is this a closed, solved case?

In this paper, we measure the radial transport of cold gas in cosmic-noon disks from the VELA zoom-in hydro-cosmological simulations \citep[][]{ceverino14, zolotov15} and compare this to analytical predictions from the VDI-driven radial mass transport models and the observational estimates of \citet{genzel23}. The simulations incorporate gravity, Eulerian gas dynamics using an adaptive mesh refinement (AMR) scheme with a maximum AMR resolution of $17.5-35 \pc$ in physical units at all times, and additional physics such as star formation, cooling, and feedback (see \se{2} for details). These simulations have been used to study many aspects of galaxy formation at cosmic noon \citep[e.g.,][]{ceverino14, zolotov15, ceverino16a, ceverino16b, inoue16, tacchella16a, tacchella16b, tomassetti16, mandelker17, dekel20, lapiner23}.

We recall that in massive halos at cosmic noon, cold gas from the cosmic web is funneled into hot halos through a few dominant filaments or streams, delivering it deep within the halo to the host galaxy \citep[][]{birnboim03, keres05, dekel06, ocvirk08, dekel09a, pichon11, vandevoort11, danovich12, danovich15, nelson13, neslon15, padnos18, mandelker18, aung19, mandelker19, mandelker20}. These streams are likely to play an additional role in driving the radial transport of cold gas in disks, either directly by their inward velocities and/or indirectly by driving disk turbulence. This is expected to introduce deviations from cylindrical symmetry, which are not captured by the VDI models. 

Investigating the radial transport of cold gas in $z \sim 0$ Milky Way (MW) mass disks using zoom-in hydro-cosmological simulations of the FIRE suite \citep[][]{hopkins14}, \citet[][]{trapp22} find that most of the freshly incoming gas piles up near the disk edge and does not contribute much to the radial mass flux within the disk. However, analyzing MW halos from the Auriga suite of simulations \citep[][]{grand17}, \citet[][]{okalidis21} show that the accreted gas fraction can be large upto beyond $70 \%$ of the disk radius, leading to increased average radial inflow speeds in this region, which then subside to a constant value of $\sim \! 2\ {\rm km/s}$ in the inner disk. Cosmic noon disks may have even higher stream penetrability, as they are fed by more intense streams \citep[][]{dekel06, dekel09a}. In fact, in an earlier observational study by \citet{martin19}, radial motions in two $z \sim 2$ Ly$\alpha$ emitting nebulae were found to be consistent with filamentary accretion-driven inflows. Here, using the VELA simulations, we measure the direct contribution of the streams to the radial transport of cold gas in simulated cosmic-noon disks. Distinguishing freshly incoming streams from the off-stream disk gas is non-trivial in an Eulerian simulation, and we resort to crude stream identification techniques (see \se{st} for details). In previous work, \citet{martin19} measured the radial inflow velocities in VELA 7 at $z \sim 2$, finding them to be consistent with the observed motions in their sample of Ly$\alpha$ emitting nebulae. However, the analysis was restricted to one snapshot only and no quantitative distinction between freshly incoming streams and off-stream material was made.

The remainder of this paper is organized as follows. In \se{2}, we describe the VELA simulations, define the main physical quantities of interest, and demonstrate our method for selecting rotation-supported disks. In \se{3}, the radial transport of cold gas in the VELA disks is discussed, and in \se{4}, these results are compared to expectations from the VDI-driven radial mass transport models. In \se{st}, freshly incoming streams of cold gas are identified, and the radial transport of streams and off-stream material are separately quantified. In \se{compare_to_obs}, the simulation results are compared to the \citet{genzel23} observations. We summarize and conclude in \se{7}.

\section{Analyzing the Simulations}
\label{sec:2}

We begin this section by briefly describing the VELA simulations (\se{sims}). Next, in \se{phyquant}, we introduce the main physical quantities of interest, particularly those relevant for characterizing radial transport. We end by demonstrating the selection of rotation-supported disks from the VELA simulation suite in \se{diskgs}.

\subsection{VELA Simulations}
\label{sec:sims}

The VELA simulation suite consists of 34 hydro-cosmological zoom-in simulations of galaxies, typically evolved to $z \sim 1$, with halo masses in the range of $\sim \! 10^{11}-10^{12} \Msun$ at $z \sim 2$ \citep[][]{ceverino14, zolotov15, ceverino16a, ceverino16b, mandelker17}. The simulations are run with the Adaptive Refinement Tree (ART) code \citep[][]{kravtsov97,kravtsov03,ceverino09}, which accurately follows the evolution of a gravitating $N$-body system and Eulerian gas dynamics using an AMR scheme. The maximum AMR resolution of the simulations is $17.5-35 \pc$ in physical units at all times. In addition to gravity and hydrodynamics, the simulations incorporate the physics of gas and metal cooling, UV-background photoionization, stochastic star formation, gas recycling, stellar winds and metal enrichment, thermal feedback from supernovae \citep[][]{ceverino10,ceverino12}, and feedback from radiation pressure \citep[][]{ceverino14}. The dark matter particles have a mass of $8.3 \times 10^{4} M_{\odot}$ and stellar particles are formed with a minimum mass of $10^{3} M_{\odot}$. For further details on the simulation method and the VELA simulation suite in general, see \citet[][]{ceverino10, ceverino14, zolotov15, mandelker17}.

\subsection{Physical Quantities}
\label{sec:phyquant}

There are 1098 snapshots in total across all 34 galaxies, equally spaced in the cosmological expansion factor, $a=(1+z)^{-1}$, with $\Delta a=0.01$. For each snapshot, the galactic center is determined iteratively using only the innermost stars \citep[stars within spheres of decreasing radii from 600 to 130 pc, initially centered at the minimum of the potential well; for more details, see][]{mandelker14}. A `cold' disk is then defined using gas with temperatures less than $1.5 \times 10^4 \Kdegree$ and stars with ages less than $100 \Myr$. The unit vector, $\hat{z}$, along the angular momentum of this cold component (in its rest frame) determines the disk plane, i.e., the plane perpendicular to $\hat{z}$. The direction of $\hat{z}$, the disk radius, $R_\rmd$, and disk height (half thickness), $H_\rmd$, are computed iteratively until they converge to within $5 \%$, following the procedure described in Appendix~B of \citet{mandelker14}. Briefly, $R_\rmd$ is defined such that it encloses $85 \%$ of the cold mass within a cylinder of radius $0.15 R_\rmv$ and height $1 \kpc$. Here, $R_\rmv$ is the virial radius of the dark matter halo hosting the galaxy, defined according to the \citet{bryan98} critical overdensity criterion. $H_\rmd$ is defined such that the disk thickness, $2 H_\rmd$, contains $85 \%$ of the cold mass within a cylinder of radius and height equal to $R_\rmd$. The angular momentum of the cold material within a cylinder of radius $R_\rmd$ and height $H_\rmd$ determines $\hat{z}$. See \citet{mandelker14} for more details.

Except when otherwise stated, here onward, we focus on the cold gas within the disk, i.e., gas with temperatures less than $1.5 \times 10^4 \Kdegree$ and residing within a cylinder of radius $R_\rmd$ and height $H_\rmd$. This is for comparison to the H$\alpha$/CO observations of \citet{genzel23}. However, we have verified that at least $80 \%$ (typically around $85-90 \%$) of the gas mass in VELA disks is in the cold phase defined by this temperature threshold and that increasing the temperature threshold to include warm or hot gas with higher temperatures has no significant impact on our results (see also Appendix~\ref{sec:A5}). The ratio of the cold gas mass to the baryonic (gas + stars) mass within the disk is defined as the cold gas fraction, $f$. In \se{st}, the cold gas within the disk is further classified into recently accreted streams versus the off-stream material, labeled as non-streams.

Two key physical quantities relevant to the study of radial transport are velocity and mass flux in the radial direction. These are determined for all cold gas as well as for streams and non-streams separately. The average radial velocity and radial mass flux as a function of the cylindrical distance from the galactic center, $r$, are defined as
\begin{equation}
    V_r = \frac{\Sigma^N_{i=1} v_{r,i}\ m_i}{M} 
    \label{eq_1}
\end{equation}
and 
\begin{equation}
    F_r = \frac{1}{\Delta r} \Sigma^N_{i=1} v_{r,i}\ m_i \,,
     \label{eq_2}
\end{equation}
respectively, where
\begin{equation}
    M = \Sigma^N_{i=1} m_i \,.
     \label{eq_3}
\end{equation}
Here, $v_{r,i}$ and $m_i$ are the radial velocity and mass of the $i^{\rm th}$ gas cell, respectively, and the sums are over cells containing all cold gas, streams, or non-streams, as relevant, in a cylindrical shell of radius $r$, width $\Delta r$, and height, $H_\rmd$. $N$ is the total number of relevant cells within this volume, and $M$ is their combined mass. Both $V_r$ and $F_r$ are always quoted as dimensionless quantities. In the case of the former, this is achieved by computing its ratio with respect to $V_{\rm rot}$, the average rotational velocity of all cold gas as a function of $r$, obtained from Equation~\ref{eq_1} by replacing $v_{r,i}$ with $v_{{\rm rot},i}$, the rotational velocity of the $i^{\rm th}$ gas cell. Similarly, the latter is divided by $\mathscr{M}/t_{\rm dyn}$, where $\mathscr{M}$ is the same as the cold gas $M$ as a function of $r$, obtained from Equation~\ref{eq_3}, and $t_{\rm dyn}=r/V_{\rm rot}$ is the local dynamical time. 

Another important physical quantity is the radial velocity dispersion of cold gas as a function of $r$, defined as 
\begin{equation}
  \sigma_r = \sqrt{\frac{\Sigma^N_{i=1} v^2_{r,i}\ m_i}{\Sigma^N_{i=1} m_i}- \left( \frac{\Sigma^N_{i=1} v_{r,i}\ m_i}{\Sigma^N_{i=1} m_i}  \right)^2}\,,
  \label{eq_4}
\end{equation}
where the sums are over all cold gas cells within a cylindrical shell of radius $r$, width $\Delta r$, and height, $H_\rmd$. Note that the second term in the expression under the square root is nothing but the cold gas $V_r$ raised to the power of $2$.

\begin{figure}
    \centering
    \includegraphics[width=0.45\textwidth]{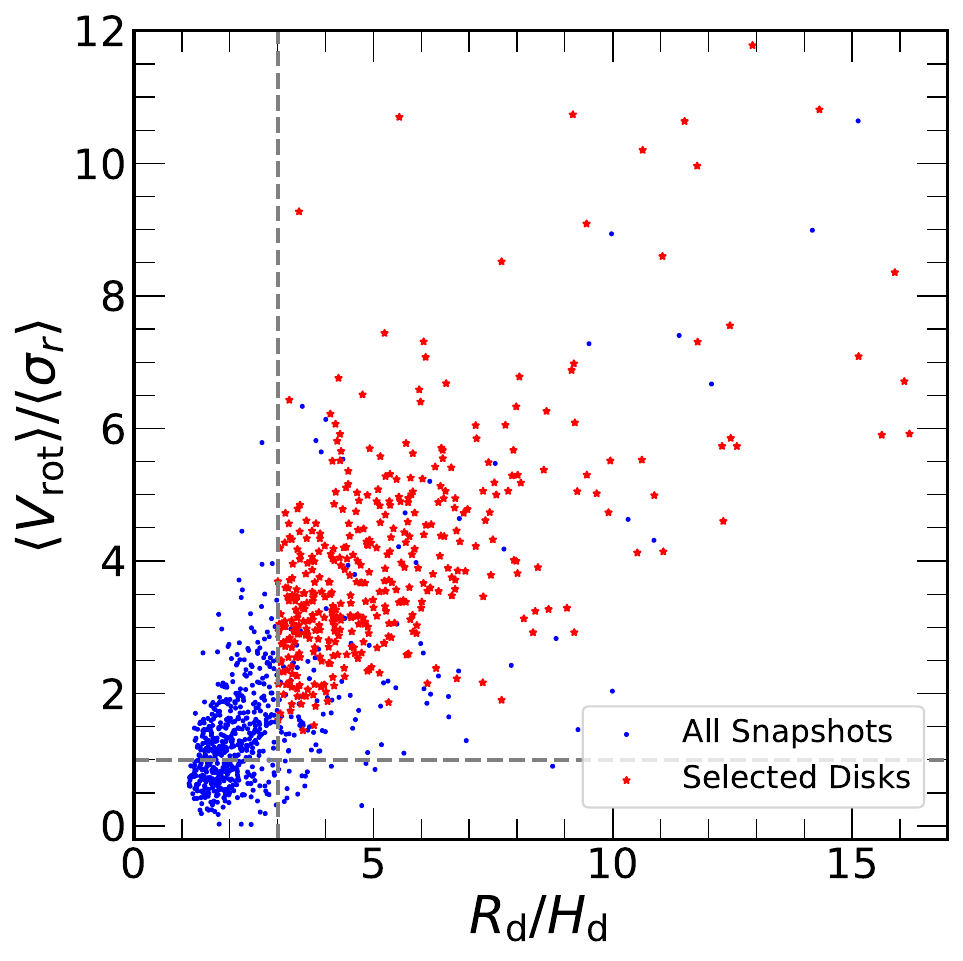}
    \caption{{\bf Disk Selection}: The ratio of the disk-averaged rotational velocity to the disk-averaged radial velocity dispersion, $\langle V_{\rm rot} \rangle/\langle \sigma_{r} \rangle$, is plotted against the ratio of the disk radius to the disk height, $R_\rmd/H_\rmd$, for all 1098 snapshots belonging to 34 VELA galaxies with the blue, round points. Of these, 425 snapshots satisfy the conditions $R_\rmd/H_\rmd>3$ and $V_{\rm rot}/\sigma_{r}>1$ at all $r$, the distance from the galactic center in cylindrical coordinates, and are selected as rotation-supported disks, highlighted by the red stars. Here, $V_{\rm rot}$ and $\sigma_{r}$ are the average rotational velocity and the radial velocity dispersion of cold gas as a function of $r$, respectively. The gray, dashed vertical and horizontal lines indicate $R_{\rmd}/H_{\rmd}=3$ and $\langle V_{\rm rot} \rangle/\langle \sigma_{r} \rangle=1$, respectively. Some of the blue points with $R_\rmd/H_\rmd>3$ and $\langle V_{\rm rot} \rangle/\langle \sigma_{r}\rangle>1$ do not satisfy the condition $V_{\rm rot}/\sigma_{r}>1$ at all $r$. As such, they are not in the disk galaxy sample.}
    \label{fig:selected_disks}
\end{figure}

By constructing ten equally spaced bins with $\Delta r=0.1 R_\rmd$, the radial profiles of $V_r$, $V_{\rm rot}$, and $\sigma_r$ are obtained over the interval $r=0$ to $R_\rmd$. These are further mass-weighted averaged over $r$ to give $\langle V_r \rangle$, the stream-, non-stream-, or disk-averaged radial velocity, as relevant, $\langle V_{\rm rot} \rangle$, the disk-averaged rotational velocity, and $\langle \sigma_r \rangle$, the disk-averaged radial velocity dispersion.

\subsection{Disk Galaxy Sample}
\label{sec:diskgs}

\begin{figure*}
    \centering
    \includegraphics[width=0.95\textwidth]{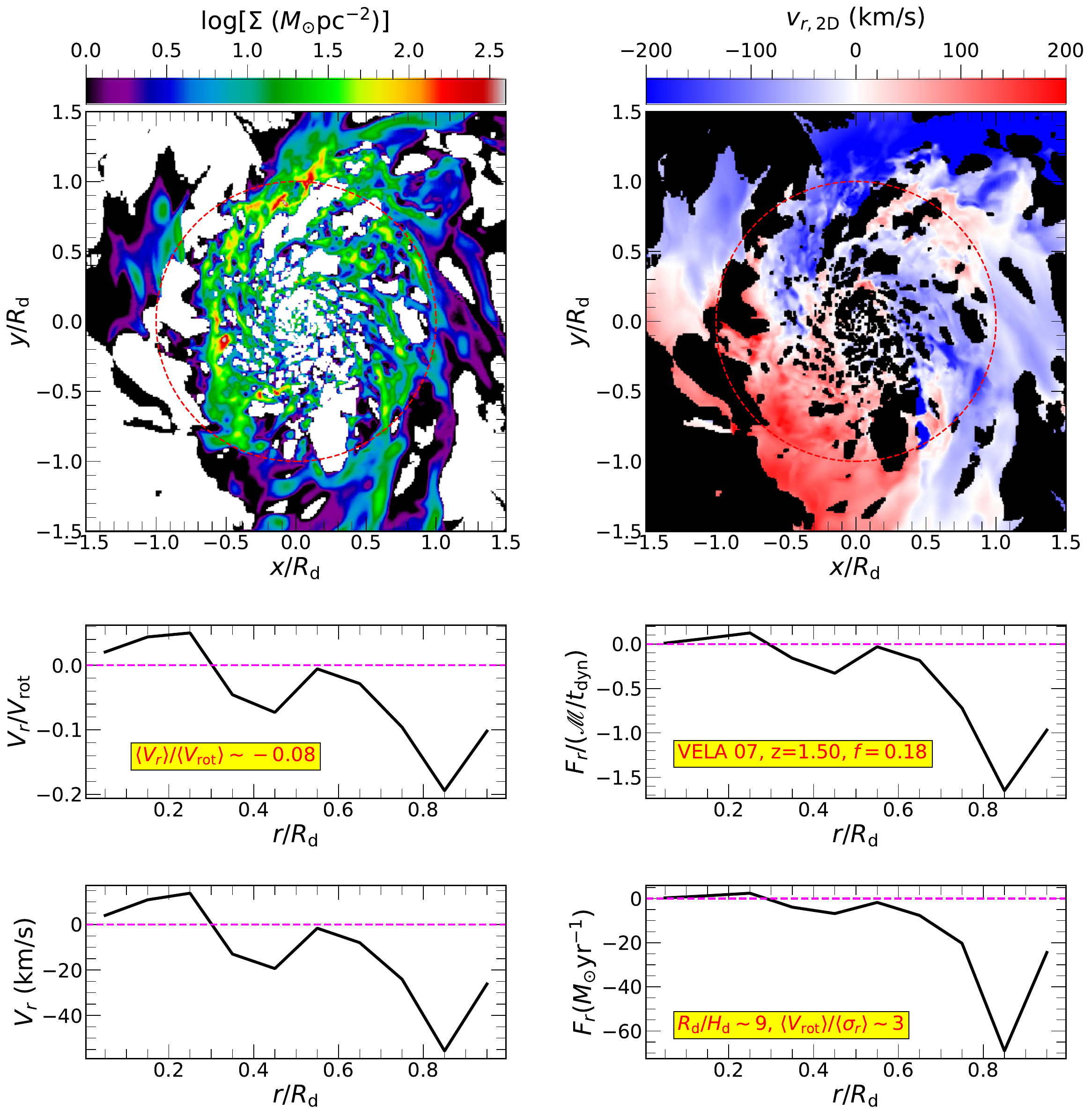}
    \caption{{\bf Radial Transport in a VELA Disk}: For the simulated galaxy VELA 7 at a redshift of $z=1.5$, the top left- and right-hand panels show face-on maps of the cold gas surface density ($\Sigma$) and the two-dimensional radial velocity ($v_{r,\rm 2D}$), respectively, within a region spanning $3\ R_\rmd \times 3\ R_\rmd$, centered on the galactic center. The disk is large ($R_\rmd=18.43\ \rm kpc$) and thin ($R_\rmd/H_\rmd \sim 9$). It is mainly supported by rotation, with $\langle V_{\rm rot} \rangle/\langle \sigma_{r} \rangle \sim 3$, and has a cold gas fraction of $f=0.18$. Regions with no cold gas are shown as white in the surface density map and black in the radial velocity map. Dashed red circles mark the radial extent of the disk. Both maps exhibit strong non-axisymmetry, largely due to incoming cold gas streams. The middle left- and right-hand panels display the average radial velocity ($V_r$, in units of $V_{\rm rot}$) and the radial mass flux ($F_r$, in units of $\mathscr{M}/t_{\rm dyn}$), respectively, as functions of $r$, normalized by $R_\rmd$, out to the disk radius. Here, $V_{\rm rot}$ and $\mathscr{M}$ are the average rotational velocity and mass of all cold gas at $r$, and $t_{\rm dyn}=r/V_{\rm rot}$ is the local dynamical time. The non-normalized $V_r$ and $F_r$ are shown in the bottom left- and right-hand panels, respectively. Beyond roughly $0.2\ R_\rmd$, both $V_r$ and $V_{\rm rot}$ are negative and tend to grow in magnitude with increasing $r$, albeit with some fluctuations, indicating a predominantly inflow-driven cold gas radial transport within the disk. The disk-averaged radial velocity is also negative and has a magnitude about $8 \%$ of the disk-averaged rotational velocity.}
    \label{fig:ex}
\end{figure*}

\begin{figure*}
    \centering
    \includegraphics[width=0.95\textwidth]{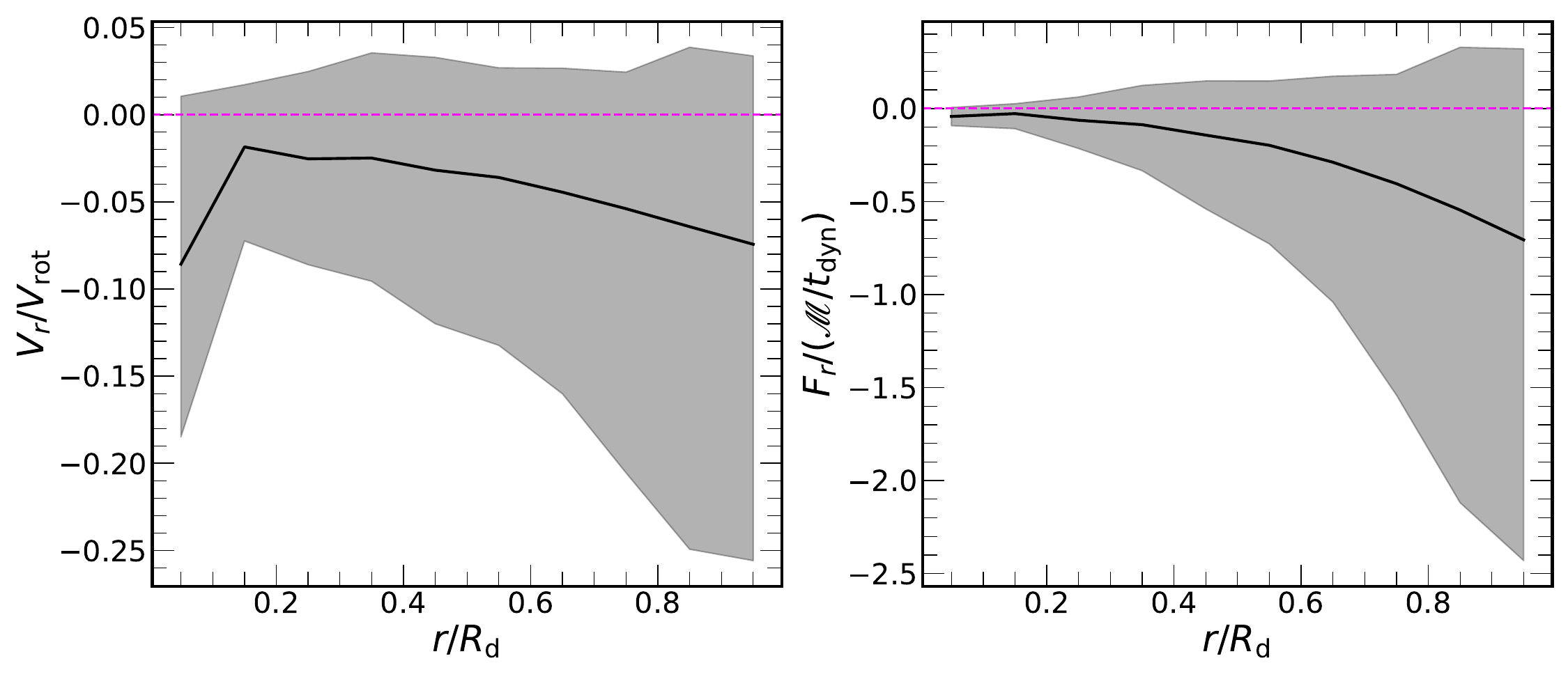}
    \caption{{\bf Trends with Distance from the Galactic Center}: The left- and right-hand panels show, respectively, the medians (solid curves) and the $16^{\rm th}$–$84^{\rm th}$ percentile ranges (shaded regions) for the average radial velocity ( $V_r$, in units of $V_{\rm rot}$), and the radial mass flux ($F_r$, in units of $\mathscr{M}/t_{\rm dyn}$), as functions of $r$ (in units of $R_\rmd$), derived from our sample of disk galaxies. The median values of both $V_r$ and $F_r$ are consistently negative and become more negative with increasing $r$, except within approximately $\sim \! 0.2\ R_\rmd$, where the trend reverses. Although the $84^{\rm th}$ percentile values of both $F_r$ and $V_r$ are positive at all radii, the magnitudes at the $16^{\rm th}$ percentile are substantially larger. Therefore, statistically speaking, the radial transport of cold gas across our sample is dominated by inflows at all radii.}
    \label{fig:stats_r}
\end{figure*}

From the 1098 snapshots, rotation-supported disks are selected by requiring that $R_\rmd/H_\rmd>3$ and $V_{\rm rot}/\sigma_{r}>1$ at all $r$. The former is a reasonable cut to ensure a disk-like geometry, while the latter affirms a rotational support that is larger than the dispersion support at all $r$. This selection is both for comparison to disk-instability-based transport models, which are valid for rotation-supported disks, and also to the \citet[][]{genzel23} sample of galaxies. The minimum $\langle V_{\rm rot}\rangle /\langle \sigma_{r} \rangle$ of our (selected) disk galaxy sample is around $1.5$, while that of the \citet[][]{genzel23} galaxies is $2$. Our sample comprises 425 snapshots, about $\sim 39 \%$ of all VELA snapshots. In Fig.~\ref{fig:selected_disks}, $\langle V_{\rm rot} \rangle/\langle \sigma_{r} \rangle$, which is a measure of the disk-averaged rotational support in the galaxy compared to the disk-averaged dispersion support, is plotted against $R_\rmd/H_\rmd$ for all VELA snapshots with the blue, round points. The red stars highlight the disk galaxy sample. The gray, dashed vertical and horizontal lines denote $R_\rmd/H_\rmd=3$ and $\langle V_{\rm rot} \rangle/\langle \sigma_{r} \rangle=1$, respectively. Note the blue points with $R_\rmd/H_\rmd>3$ and $\langle V_{\rm rot} \rangle/\langle \sigma_{r}\rangle>1$ that are not in the disk galaxy sample, as they do not satisfy the condition $V_{\rm rot}/\sigma_{r}>1$ at all $r$. However, we have verified that including these snapshots in the disk sample makes no significant difference to our results.

\section{Cold Gas Radial Velocity and Mass Flux}
\label{sec:3}

In this section, we begin describing the radial transport of cold gas in the VELA simulations, focusing on the radial velocity and mass flux (see \se{phyquant}). In \se{ex}, we present an illustrative example. General trends with respect to the distance from the galactic center, cold gas fraction, and redshift are highlighted in \se{stats}.

\subsection{An Illustrative Example}
\label{sec:ex}

\begin{figure*}
    \centering
    \includegraphics[width=0.95\textwidth]{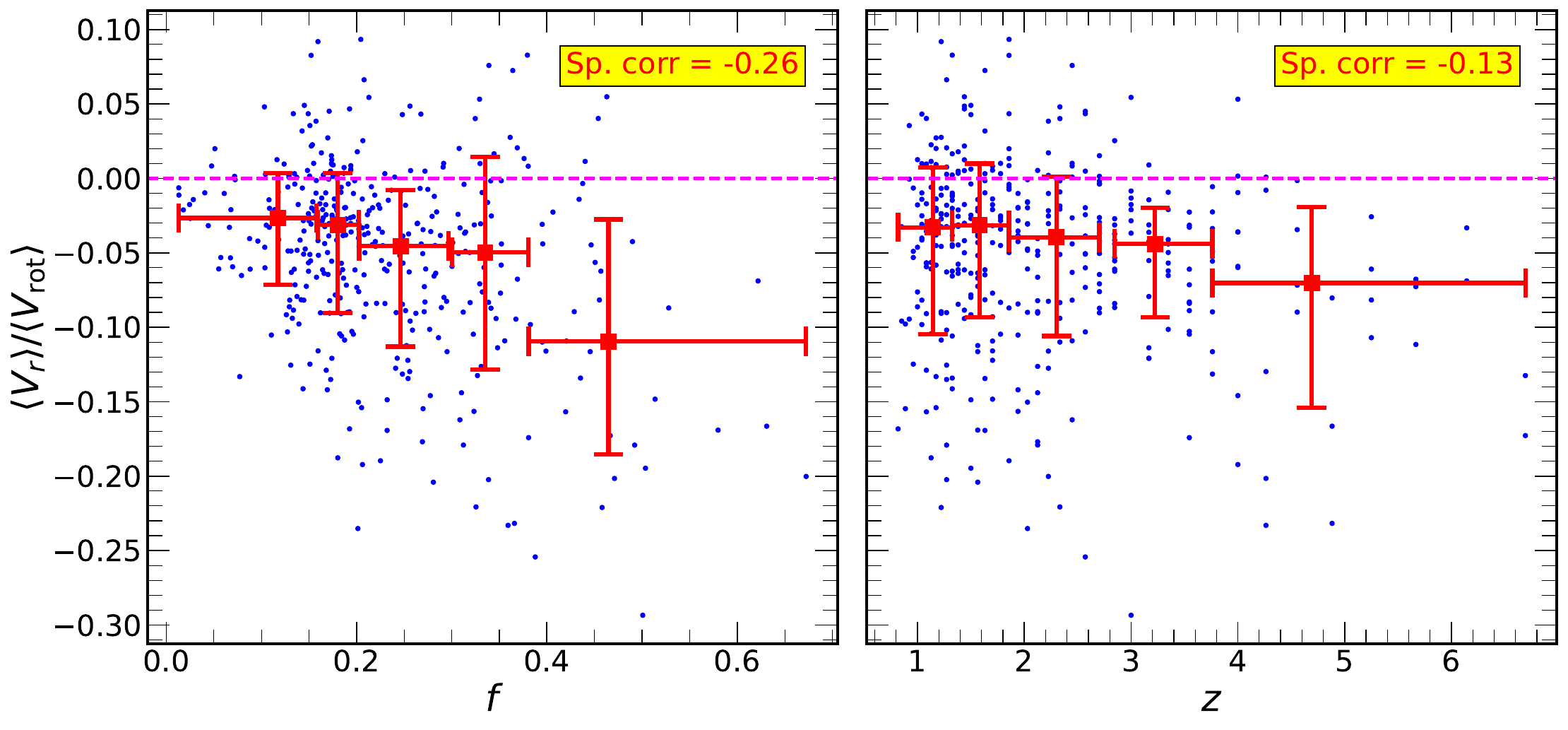}
    \caption{{\bf Trends with Cold Gas Fraction and Redshift}: The left- and right-hand panels show the ratio of the disk-averaged radial velocity to the disk-averaged rotational velocity, $\langle V_r \rangle/\langle V_{\rm rot} \rangle$, plotted as a function of the cold gas fraction ($f$) and redshift ($z$), respectively, using blue circles. Red squares represent the median values within bins of $f$ and $z$ (chosen to ensure adequate snapshot counts), while the vertical error bars correspond to the $16^{\rm th}$–$84^{\rm th}$ percentile ranges. The horizontal error bars indicate the bin widths, and the squares are placed at the mean values of $f$ and $z$ for each bin. For the majority of VELA disks ($82 \%$), $\langle V_r \rangle$ is negative (note that $\langle V_{\rm rot} \rangle$ is strictly positive), underscoring the dominance of cold gas radial inflows over outflows in our disk galaxy sample. On average, $\langle V_r \rangle/\langle V_{\rm rot} \rangle$ becomes more negative with increasing $f$ and $z$, suggesting that galaxies with higher cold gas content or located at higher redshifts typically exhibit stronger radial inflows. Although both correlations are relatively weak, the trend with cold gas fraction is stronger than that with redshift, as reflected in the Spearman rank correlation coefficients of $-0.26$ and $-0.13$, respectively.}
    \label{fig:stats_fg_z}
\end{figure*}

In Fig.~\ref{fig:ex}, we demonstrate the radial transport of cold gas in one of the VELA galaxies, VELA 7, at a redshift of $1.5$. At this epoch, the disk is quite large and thin ($R_\rmd=18.43\ \rm kpc$, $R_\rmd/H_\rmd \sim 9$). It has a dominant rotational support with $\langle V_{\rm rot} \rangle/\langle \sigma_r \rangle \sim 3$ and a gas fraction of $0.18$. 

The top left- and right-hand panels, respectively, show the maps of face-on projected surface density, $\Sigma$, and radial velocity, $v_{r, \rm 2D}$, for the cold gas in a region of dimensions $3\ R_\rmd \times 3\ R_\rmd$, centered on the galactic center. To obtain these maps, all relevant gas data (e.g., density, velocity, temperature) is first interpolated onto a uniform 3D grid with cell size, $s= 0.01\ R_\rmd$ along each direction. Then, for a particular ($x$, $y$), $\Sigma$ is obtained from the interpolated density field, $\rho$, by summing $\rho \times s$ over all cold gas cells between $z=-H_\rmd$ to $H_\rmd$, where $z=0$ represents the disk mid-plane. Similarly, $v_{r, \rm 2D}$ at a particular ($x$, $y$) is determined as the mass-weighted average of the interpolated radial velocity field, $v_r$, between $z=-H_\rmd$ to $H_\rmd$, only considering cold gas cells. Regions devoid of cold gas are shown with white and black patches in the surface density and radial velocity maps, respectively. The red, dashed circles indicate the radial extent of the disk. 

We observe that both the radial velocity and surface density maps are highly non-axisymmetric, largely representing incoming streams of cold gas. At least two distinct streams can be visually identified in this galaxy, one inflowing into the disk (entering within $R_\rmd$) from the top and the other from the bottom-right of the maps (in the $v_{r, \rm 2D}$ map, blue indicates inflows). In addition, most of the red, outflowing region in the bottom-left portion of the maps may be interpreted as a part of the former stream, en route to the apocenter after having crossed the pericenter. Streams are discussed in detail in \se{st}. For now, just by looking at the maps, it suffices to say that a significant fraction of the cold gas mass within the disk is in streams with large radial velocities (positive or negative). Thus, they are likely to play a prominent role in the story of radial transport.

In the middle left- and right-hand panels, we, respectively, show the cold gas $V_r$ (in units of $V_{\rm rot}$) and $F_r$ (in units of $\mathscr{M}/t_{\rm dyn}$) as functions of $r$ (in units of $R_\rmd$) out to the disk radius. The non-normalized $V_r$ and $F_r$ are depicted in the bottom left- and right-hand panels, respectively. Except within $\sim \! 0.2\ R_\rmd$, $V_r$ and $F_r$ are both negative and generally increase in magnitude with an increase in $r$, though there are some oscillations. $V_r$ is most negative near $r=0.8\ R_\rmd$, attaining a value of $-0.2\ V_{\rm rot}$. $F_r$ is also most negative at the same radius. The disk-averaged radial velocity, $\langle V_r \rangle$, is also negative and about $8 \%$ of the disk-averaged rotational velocity, $\langle V_{\rm rot} \rangle$, in magnitude. Therefore, the radial transport of cold gas in the disk is inflow-dominated.

\subsection{General Trends}
\label{sec:stats}

Having discussed a particular case, we now describe some general trends. In Fig.~\ref{fig:stats_r}, the left- and right-hand panels, respectively, show the medians (solid lines) and $16^{\rm th}-84^{\rm th}$ percentiles (envelopes) of cold gas $V_r$ (in units of $V_{\rm rot}$) and $F_r$ (in units of $\mathscr{M}/t_{\rm dyn}$) as functions of $r$ (in units of $R_\rmd$) obtained from our disk galaxy sample. The median $V_r$ and $F_r$ are always negative and increase in magnitude with increasing $r$, except within $\sim \! 0.2\ R_\rmd$, where the trends are reversed. While both $F_r$ and $V_r$ are positive at the $84^{\rm th}$ percentile for all $r$, their magnitudes at the $16^{\rm th}$ percentile are much larger. For example, at $r=R_\rmd$, $V_r$ at the $16^{\rm th}$ percentile is about $-0.25\ V_{\rm rot}$, while that at the $84^{\rm th}$ percentile is only $\sim \! 0.04\ V_{\rm rot}$. Similarly, at the same radius, $F_r$ at the $16^{\rm th}$ percentile is about five times larger in magnitude than that at the $84^{\rm th}$ percentile. Therefore, in a statistical sense, the radial transport of cold gas in our disk galaxy sample is inflow-dominated at all radii, i.e., the inflowing flux is larger in magnitude than the outflowing flux. This is also shown explicitly in the right-hand panels of Figure~\ref{fig:stats_pos_neg}.  

In Fig.~\ref{fig:stats_fg_z}, the ratio of the disk-averaged radial velocity to the disk-averaged rotational velocity, $\langle V_r \rangle/\langle V_{\rm rot} \rangle$, is plotted against $f$ and $z$ in the left- and right-hand panels, respectively, with the blue, round points. The red squares and the vertical error bars denote the medians and $16^{\rm th}-84^{\rm th}$ percentiles, respectively, in $f$ and $z$ bins of varying widths (to include a sensible number of snapshots), as indicated by the horizontal error bars, with the locations of the squares along the $f$- and $z$-axes being at the bin-averages. We find that for most of the VELA disks ($82 \%$), $\langle V_r \rangle$ is negative, again highlighting the inflow-dominated nature of cold gas radial transport in our disk galaxy sample (note that $\langle V_{\rm rot} \rangle$ is always positive). Also, on average, $\langle V_r \rangle/\langle V_{\rm rot} \rangle$ decreases (increases in magnitude) with an increase in both $f$ and $z$, indicating that galaxies with larger cold gas fractions or at higher redshifts have higher levels of disk-averaged radial inflow in general. Both trends are relatively weak, though the one with cold gas fraction is more pronounced than that with redshift, the Spearman's rank correlation coefficients in the two cases being $-0.26$ and $-0.13$, respectively. 

If the in-disk radial motions are driven by internal disk instabilities, a strong correlation is expected with gas fraction (and redshift, as gas fractions increase with increasing redshift). That we only see a weak correlation hints against this scenario, which is explored in more details in the next section. Trend in the radial motion with total cold gas mass in explored in Appendix~D.

\section{Disk Instability Based Models of Radial Transport}
\label{sec:4}

Disk instability can induce gravitational torques that lead to angular momentum transport outwards and mass transport inwards. Below, we first summarize five different radial transport models based on disk instability (\se{viscous_transport}-\se{clump_migration}) and then compare their predictions for the inflow or inward migration velocity with the results from the VELA disk galaxy sample (\se{compare_to_theory}). 

\subsection{Viscous Disk Transport}
\label{sec:viscous_transport}

\citet[][KB10 hereafter]{krumholz10} and \citet[][K18 hereafter]{krumholz18} assume an axisymmetric disk of gas and stars that is always kept in a state of marginal instability. If the disk becomes gravitationally unstable, turbulence in the gas acts like viscosity, breaking axisymmetry and inducing torques that drive gas radially inwards until marginal instability is restored. In addition, the gas disk is assumed to be in energy equilibrium, implying that the energy lost due to dissipation of turbulence is balanced by the energy gained from star formation feedback and input of gravitational energy from non-axisymmetric torques. Under these conditions, a steady-state solution for the gas mass inflow rate is given by 
\begin{equation}
    \dot{M}_{\rm in} = \frac{1}{(1-\beta)}\frac{\mathscr{L} 2 \pi r^2}{V^2} \left( 1-\sigma_{\rm sf}/\sigma \right) \,.
    \label{M_dot}
\end{equation}
Here, $\mathscr{L}$ is the rate at which turbulent kinetic energy per unit area is dissipated, $V$ is the circular velocity at radius, $r$, $\beta = d {\rm ln} V/ d {\rm ln} r$, $\sigma$ is the $1$D gas velocity dispersion, assumed to be isotropic, and $\sigma_{\rm sf}$ is the value of $\sigma$ that can be maintained with star formation feedback alone without any additional energy input from radial transport of gas ($\dot{M}_{\rm in}=0$). Furthermore, $\dot{M}_{\rm in}= 2 \pi r \Sigma V_{\rm in}$, where $\Sigma$ and  $V_{\rm in}$ are the gas surface density and inflow velocity, respectively. Therefore, 
\begin{equation}
    V_{\rm in} = \frac{1}{(1-\beta)} \frac{\mathscr{L} r}{V^2 \Sigma} \left( 1-\sigma_{\rm sf}/\sigma \right) \,.
    \label{V_mig}
\end{equation}

Assuming the outer scale of turbulence to be of the order of the gas scale height, the turbulence decay rate per unit area in \citetalias{krumholz10} is given by 
\begin{equation}
\mathscr{L}= \frac{\eta \Sigma \sigma^{2}}{t_{\rm cross}} \,,
\end{equation}
where $t_{\rm cross}=h/\sigma$ is the vertical crossing time, $h$ being the gas scale height at $r$. Further, assuming $V/\sigma \sim r/h$, $t_{\rm cross} = 1/\Omega$, where $\Omega=V/r$ is the angular velocity. A value of $3/2$ is adopted for $\eta$, which corresponds to all the turbulent kinetic energy being radiated away every vertical crossing time. In addition, $\sigma \gg \sigma_{\rm sf}$ is assumed, so the contribution to turbulent energy from star formation feedback is negligible. Substituting the above in Equation \ref{V_mig} and assuming a flat rotation curve ($\beta=0$) gives 
\begin{equation}
    V_{\rm in}^{\rm KB10}/V \simeq 1.5 \left( V/\sigma \right)^{-2} \,.
    \label{kb_10}
\end{equation}

In \citetalias{krumholz18}, the expression for the turbulence decay rate per unit area is
\begin{equation}
\mathscr{L}= \frac{\eta \Sigma (\sigma^2 - \sigma_{\rm th}^{2})}{t_{\rm cross}} \,,
\label{L}
\end{equation}
where $\sqrt{\sigma^2 - \sigma^2_{\rm th}}$ is the $1$D turbulent velocity dispersion of gas, $\sigma_{\rm th}$ being the $1$D thermal velocity dispersion, typically much smaller than $\sigma$, and $t_{\rm cross}=h/\sqrt{\sigma^2 - \sigma^2_{\rm th}}$. The gas scale height is estimated following \citet[][]{forbes12} as 
\begin{equation}
    h = \frac{\sigma^2}{\pi G \Sigma ( 1 + Q_{\rm g}/ Q_{*} )} \,,
    \label{h}
\end{equation}
where $G$ is the universal gravitational constant, $Q_{g}=\kappa \sigma/(\pi G \Sigma)$, and similarly for $Q_{*}$. Here, $\kappa=\sqrt{2(1+\beta)}\Omega$ is the epicyclic frequency. Therefore,
\begin{equation}
    t_{\rm cross} = \frac{1}{\sqrt{2 (1+\beta)}} \frac{1}{\phi_{Q} f_{g,Q} \phi_{\rm nt}^{1/2}}  \frac{Q}{\Omega} \,,
    \label{t_cross}
\end{equation}
where $Q$ is the two-component Toomre-Q parameter \citep[see][]{romeo13}, expressed as $Q=f_{g,Q} Q_{g}$, where $f_{g,Q}$ depends on the gas and stellar surface densities and velocity dispersions \citepalias[see Equation~9 in][]{krumholz18}, $\phi_Q=1+Q_{g}/Q_{*}$, and $\phi_{\rm nt}= 1- \sigma_{\rm th}^2/\sigma^2$. Using Equations~\ref{L} and \ref{t_cross},
\begin{equation}
   \mathscr{L} = \sqrt{2 (1+ \beta)} \eta \phi_{Q} f_{g,Q} \phi_{\rm nt}^{3/2} \frac{ \Sigma \sigma^{2} \Omega}{Q} \,.
   \label{L_new}
\end{equation}
Substituting the expression for $\mathscr{L}$ from Equation~\ref{L_new} in Equation~\ref{V_mig} and adopting the fiducial values for $\beta$, $f_{g,Q}$, $\phi_Q$, $\phi_{\rm nt}$, and $\eta$ as in \citetalias{krumholz18}, namely, $\beta=0$, $f_{g,Q}=0.5$, $\phi_Q=2$, $\phi_{\rm nt}=1$, and $\eta=1.5$ gives
\begin{equation}
    V_{\rm in}^{\rm K18}/V = \frac{2.12}{Q} \left( V/\sigma \right)^{-2}  \left( 1- \sigma_{\rm sf}/\sigma \right) \,.
\end{equation}
Additionally, if $\sigma \gg \sigma_{\rm sf}$ ($\sigma_{\rm sf} \simeq 10\ {\rm km/s}$, see \citetalias{krumholz18}), and the contribution to turbulent energy from star formation feedback can be neglected, the above equation can be further simplified to give
\begin{equation}
    V_{\rm in}^{\rm K18}/V \simeq \frac{2.12}{Q} \left( V/\sigma \right)^{-2} \,.
    \label{k18}
\end{equation}

We note that the difference in the \citetalias{krumholz10} and \citetalias{krumholz18} expressions for $V_{\rm in}$ (Equations~\ref{kb_10} and \ref{k18}) arises from the different approximations for the gas scale height, which lead to a $Q$ dependence in the scale height crossing time for the \citetalias{krumholz18} model as opposed to no $Q$ dependence in the \citetalias{krumholz10} model. If the \citetalias{krumholz18} approximation for the gas scale height (Equation~\ref{h}) were to be also adopted for the \citetalias{krumholz10} model, then the two models would yield identical expressions for $V_{\rm in}$. For more details on these models, see \citetalias{krumholz10} and \citetalias{krumholz18}.

\subsection{Ring Migration}
\label{sec:ring_migration}

In this subsection and the next, we assume an axisymmetric gas disk maintained at marginal instability, with average surface density, $\Sigma_\rmd$, disk radius, $R_\rmd$, and a flat rotation curve with circular velocity,
\begin{equation}
    V=\sqrt{GM_{\rm tot}/R_\rmd} \,,
    \label{V}
\end{equation}
where $M_{\rm tot}$ is the total mass (including the gas disk, stars, and dark matter) within $R_\rmd$. The disk mass within $R_\rmd$ is given by
\begin{equation}
    M_\rmd = \pi R_\rmd^2 \Sigma_\rmd \,.
    \label{M_d}
\end{equation}
The average Toomre-Q parameter for the disk can be expressed as
\begin{equation}
    Q=\frac{\sqrt{2} \Omega \sigma}{\pi G \Sigma_d} \simeq \sqrt{2} \delta_\rmd^{-1} \sigma/V \,,
    \label{Q}
\end{equation}
where $\sigma$ is the average $1$D gas velocity dispersion, assumed to be isotropic, the angular velocity is approximated as 
\begin{equation}
    \Omega=V/R_d \,,
    \label{omega}
\end{equation} 
and $\delta_\rmd$ is defined as $M_\rmd/M_{\rm tot}$. For marginal instability, $Q$ is of order unity, with an exact value that depends on the disk thickness.

In \citet[][D20 hereafter]{dekel20}, it is found that a long-lived ring of gas can form in a galaxy exhibiting a massive bulge. Once formed, a ring at radius, $r$ of width, $\Delta r$ migrates inwards due to non-axisymmetric torques from the rest of the disk on a timescale,
\begin{equation}
    t_{\rm in} \simeq \frac{1}{2 \pi} \frac{\Delta_r \eta_r}{mA_m^2} \delta_\rmd^{-3} \frac{2 \pi r}{V} \,, 
    \label{d20_t}
\end{equation}
which is calculated assuming a tightly-wound spiral structure with $m$ arms \citepalias[see Equation~23 in][]{dekel20}. Here, $\eta_r=\Delta r/r$ is the relative width of the ring, $\Delta_r = \Sigma_r/\Sigma_\rmd$ is the ring contrast with respect to the disk, $\Sigma_r$ being the average ring density, and $A_m$ is the fraction of the disk density that is in the spiral arm perturbation. Assuming $t_{\rm in}=r/V_{\rm in}$ and replacing $\delta_\rmd$ using Equation~\ref{Q}, Equation~\ref{d20_t} can be simplified to give
\begin{equation}
    V^{\rm D20}_{\rm in}/V \simeq 2.83 Q^{-3} (V/\sigma)^{-3} \,,
\end{equation}
where we have adopted the fiducial values for $m$, $\Delta_r$, $A_m$, and $\eta_r$ from D20, namely, $m=2$, $\Delta_r=1$, $A_m=0.5$, and $\eta_r=0.5$. For more details on this model, see \citetalias{dekel20}.

\subsection{Clump Migration}
\label{sec:clump_migration}

\begin{figure*}
    \centering
    \includegraphics[width=0.95\textwidth]{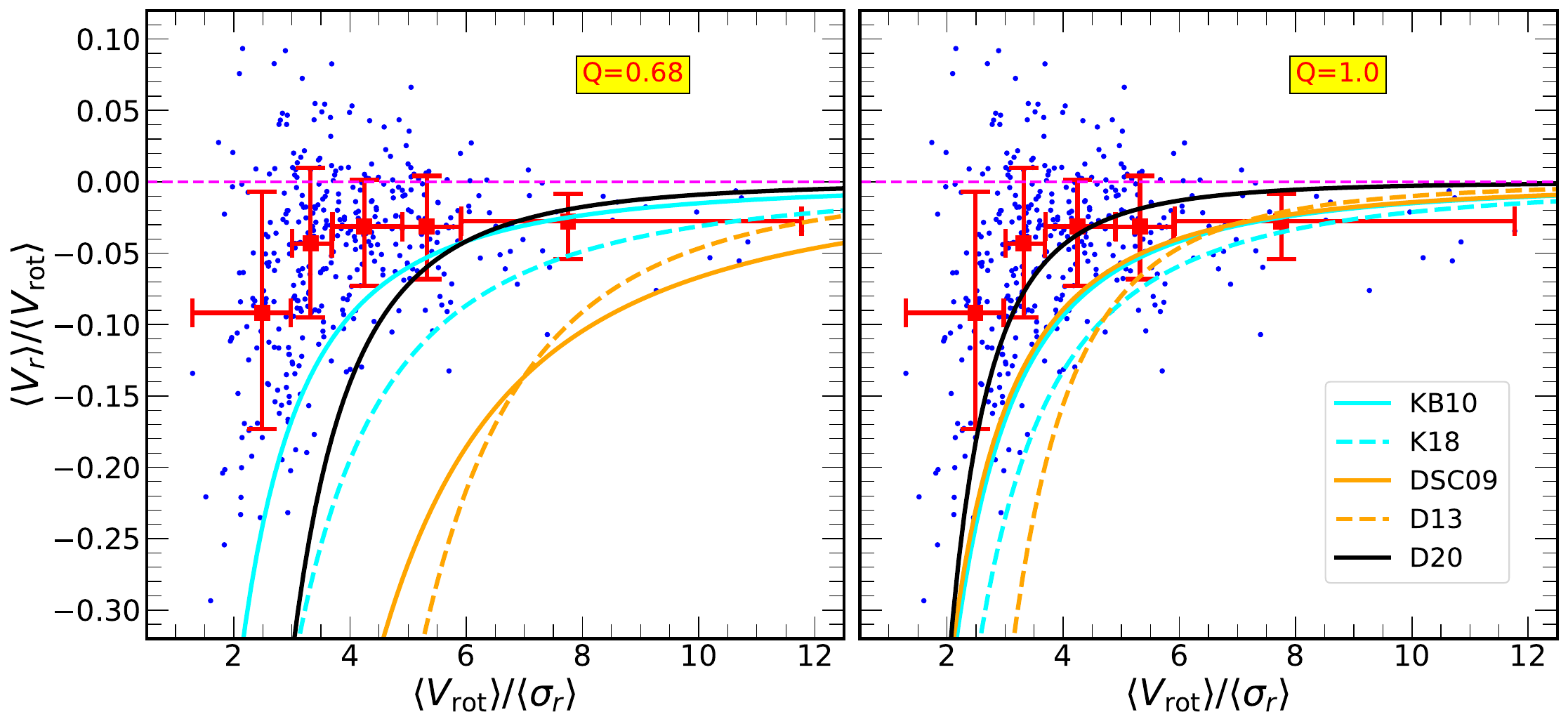}
    \caption{{\bf Comparison of Disk Instability Based Radial Transport Models with Simulations}: $\langle V_r \rangle/\langle V_{\rm rot} \rangle$ versus $\langle V_{\rm rot} \rangle/\langle \sigma_r \rangle$ is plotted for each snapshot in our disk galaxy sample, using blue circles. Red squares represent the medians within bins of $\langle V_{\rm rot} \rangle/\langle \sigma_r \rangle$ of varying widths, selected to ensure a reasonable number of snapshots per bin. Vertical error bars indicate the $16^{\rm th}$–$84^{\rm th}$ percentile ranges, and horizontal error bars show the bin widths. The red squares are positioned at the average $\langle V_{\rm rot} \rangle/\langle \sigma_r \rangle$ for each bin. Overlaid curves correspond to predictions from disk-instability-based models of radial inflow, as labeled. The left-hand panel adopts $Q = 0.68$ \citep[marginal instability for thick disks;][]{goldreich65}, and the right-hand panel uses $Q = 1$ (marginal instability for thin disks). The \citetalias{dekel20} model describes inflow via ring migration, while the \citetalias{dekel09} and \citetalias{dekel13} models focus on clump migration. The \citetalias{krumholz10} and \citetalias{krumholz18} models consider global gas transport in viscous disks and are likely the most applicable for comparison with the analysis of the VELA disks presented in this paper. It is worth noting that for most VELA disks, these two models tend to overestimate the disk-averaged radial inflow, regardless of whether $Q = 0.68$ or $Q = 1$ is assumed. Furthermore, all five models consistently predict radial inflows, whereas about $18 \%$ of the VELA disks show disk-averaged outflows.}
    \label{fig:compare_th}
\end{figure*}

Due to gravitational instability, gas-rich disks can fragment into giant clumps. As shown in \citet[][DSC09 hereafter]{dekel09}, torques from the off-clump disk mass and clump-clump encounters can cause the clumps to migrate inwards on a timescale, 
\begin{equation}
t_{\rm in} \simeq 1.4\ Q^{2} \delta_\rmd^{-2} R_\rmd/V \,.
\label{t_mig}
\end{equation}
Assuming $t_{\rm in}= R_\rmd/V_{\rm in}$ and replacing $\delta_\rmd$ using Equation~\ref{Q}, Equation~\ref{t_mig} can be simplified to give
\begin{equation}
    V^{\rm DSC}_{\rm in}/V \simeq 1.43\ Q^{-4} (V/\sigma)^{-2} \,.
\end{equation}  
For more details on this model, see \citetalias{dekel09}.

Inward clump migration can also occur due to dynamical friction on a clump from the off-clump disk mass \citep[][D13 hereafter]{dekel13}. According to Chandrasekhar's dynamical friction formula, the deceleration due to dynamical friction on a clump of mass, $M_\rmc$ moving in a circular orbit with velocity, $V$ is given by
\begin{equation}
    \dot{V} = - \frac{4 \pi G^2 {\rm ln} \Lambda M_\rmc \rho_\rmd}{V^2} \,,
    \label{chandra}
\end{equation}
where ${\rm ln} \Lambda$ is the Coulomb logarithm and $\rho_\rmd$ is the off-clump disk density. Following DSC09, and using Equations~\ref{V}, \ref{M_d}, and \ref{omega}, the pre-collapse radius of a typical clump is 
\begin{equation}
    R_\rmc \simeq \frac{\pi^2 G \Sigma_\rmd}{4 \Omega^2} = \frac{\pi}{4} \delta_\rmd R_d \,,
\end{equation}
and the clump mass is 
\begin{equation}
  M_\rmc \simeq \pi R_\rmc^2 \Sigma_\rmd  = \frac{\pi^2}{16} \delta_\rmd^2 M_\rmd \,.
\end{equation}
Assuming $H_\rmd/R_\rmd \sim \sigma/V$, where $H_\rmd$ is the disk height, and using the definition of $Q$ as in Equation~\ref{Q}
\begin{equation}
H_\rmd \simeq \frac{Q}{\sqrt{2}} \delta_\rmd R_\rmd \,.
\end{equation}
Therefore, 
\begin{equation}
\rho_\rmd \simeq \frac{M_\rmd}{2 \pi R_\rmd^2 H_\rmd} = \sqrt{2} Q^{-1} \frac{M_{\rm tot}}{2 \pi R_\rmd^3} \,.
\end{equation}
Substituting $\rho_\rmd$ and $M_\rmc$ in Equation~\ref{chandra} gives
\begin{equation}
    \dot{V} \simeq - \frac{\sqrt{2} \pi^2  V^{2} {\rm ln} \Lambda}{8 R_\rmd} Q^{-1} \delta_\rmd^{3} \,,
\end{equation}
such that the clump migration time,
\begin{equation}
    t_{\rm in} \simeq \frac{V}{|\dot{V}|} \simeq \frac{8 R_\rmd}{\sqrt{2} \pi^2 V {\rm ln} \Lambda} Q \delta_\rmd^{-3} \,.
\end{equation}
Assuming $t_{\rm in}= R_\rmd/V_{\rm in}$, the inward migration velocity of clumps due to dynamical friction is given by
\begin{equation}
    V^{\rm D13}_{\rm in}/V \simeq 9.98 Q^{-4} (V/\sigma)^{-3} \,,
\end{equation}
where following \citetalias{dekel13}, we adopt ${\rm ln} \Lambda=2$, and Equation~\ref{Q} is used to replace $\delta_\rmd$. For more details on this model, see \citetalias{dekel13}.


\begin{figure}
    \centering
    \includegraphics[width=0.45\textwidth]{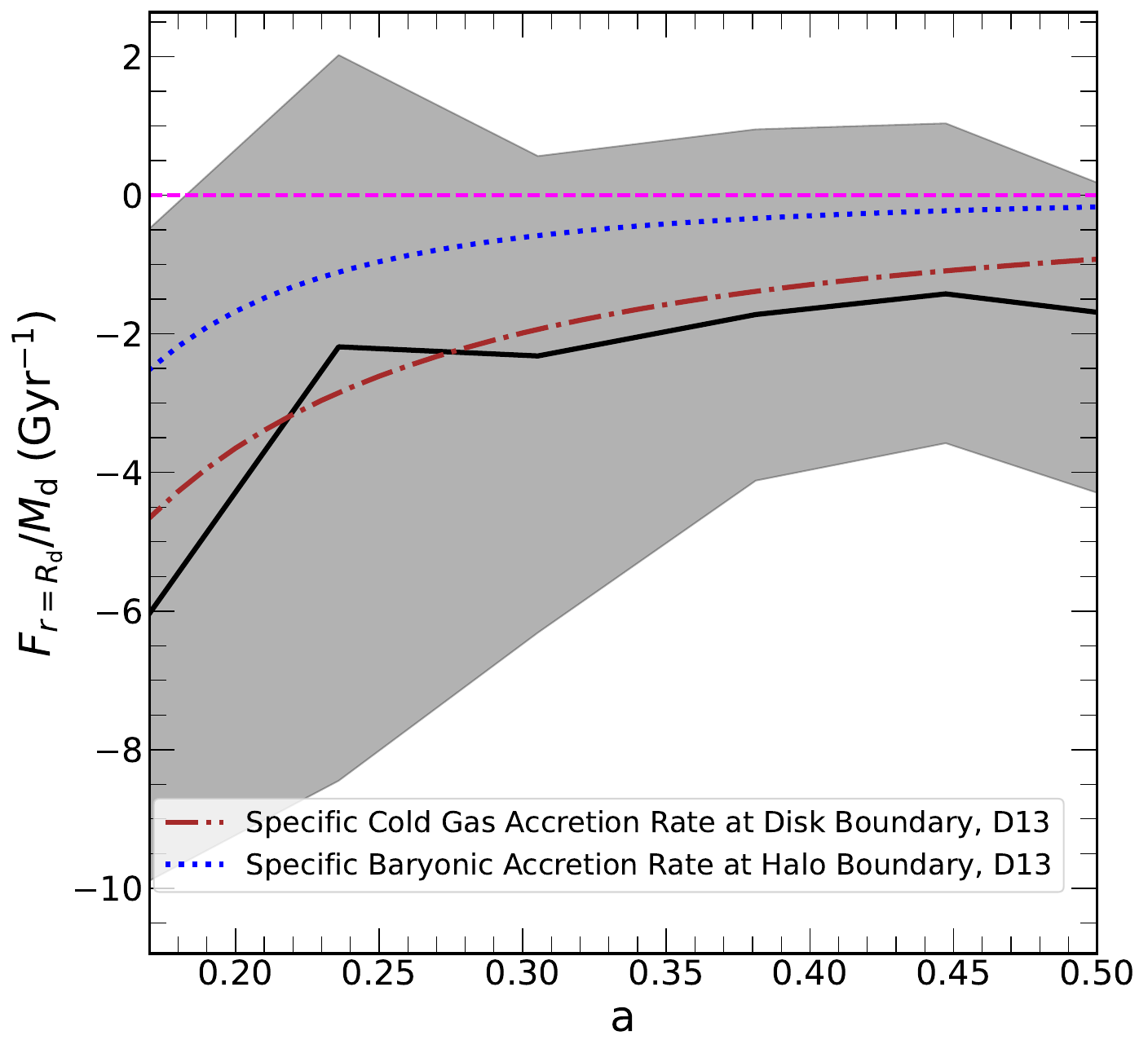}
    \caption{{\bf Cold Gas Radial Mass Flux vs Bathtub Model Prediction:} For the VELA disk galaxy sample, the cold gas radial flux mass at $R_\rmd$, $F_{r=R_\rmd}$, divided by the total cold gas mass within $R_\rmd$, $M_{\rm d}$, is plotted as a function of the scale factor, $a$, with the solid black line and the envelope indicating the median and $16^{\rm th}-84^{\rm th}$ percentile variation over $a$ bins of width $0.07$. The brown, dot-dashed line highlights the specific accretion rate of cold gas at the disk edge, predicted by a steady-state bathtub model, ignoring any disk-instability driven mass inflows through the disk onto the central bulge or feedback-driven outflows \citepalias[see][]{dekel13}. For comparison, the cosmological specific accretion rate of baryons onto the halo, also derived in \citetalias[see][]{dekel13}, is shown with the blue, dotted line. The good agreement between the black and the brown curves highlights the importance of accreted cold gas in explaining the dynamics of VELA disks.}
    \label{fig:bathtub}
\end{figure}

\begin{figure*}
    \centering
    \includegraphics[width=0.95\textwidth]{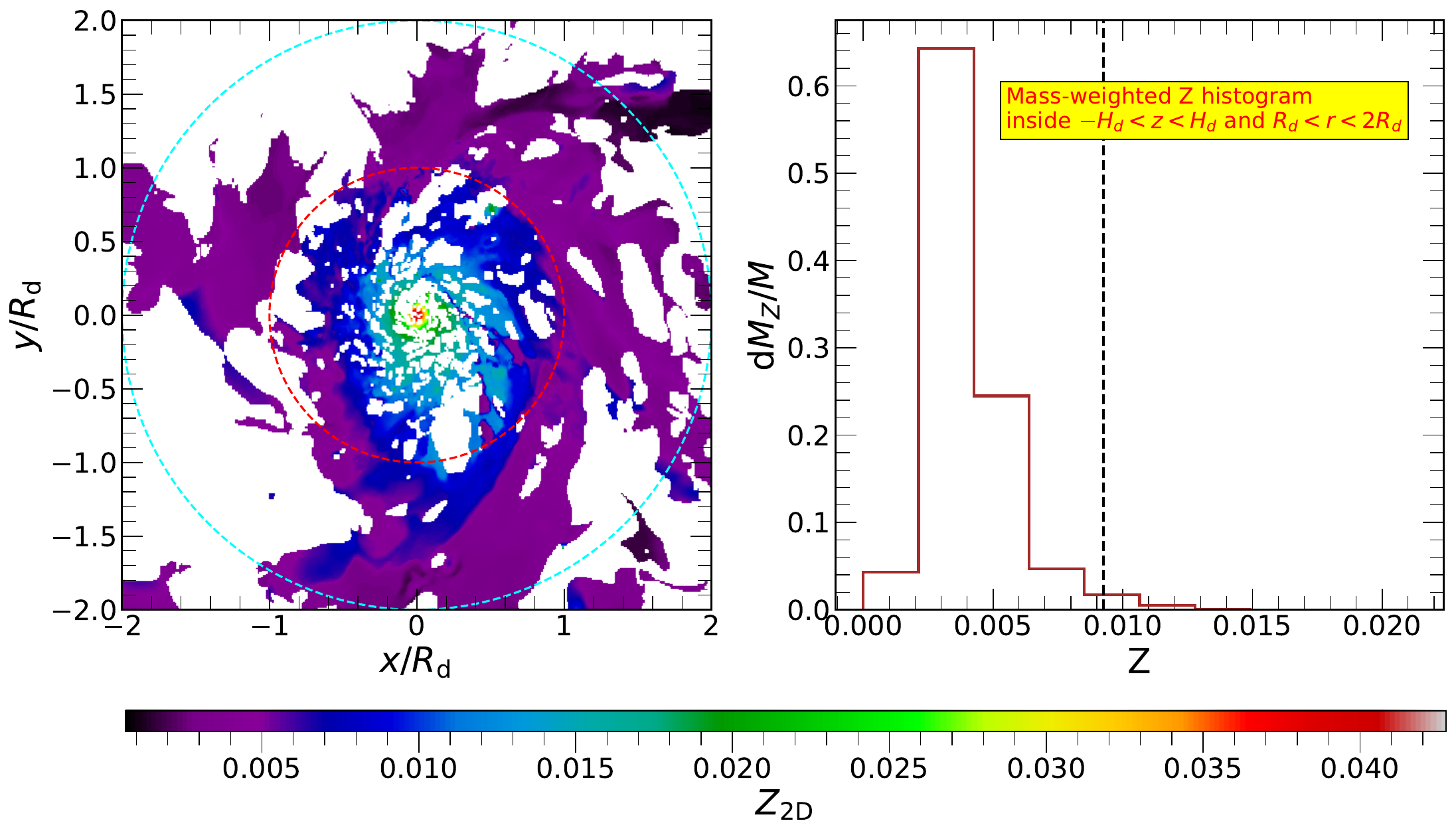}
    \caption{{\bf Metallicity-Based Selection of Streams}: The left-hand panel presents the two-dimensional metallicity distribution ($Z_{\rm 2D}$) for the VELA 7 galaxy at a redshift of $z=1.5$, shown in a face-on orientation spanning $4 R_\rmd \times 4 R_\rmd$ and centered on the galactic center. This map is derived by computing the mass-weighted average of the cold gas metallicities ($Z$) along the $z$-axis within the vertical range of $-H_\rmd$ to $H_\rmd$. Overlaid on the map are dashed red and cyan circles representing radii of $R_\rmd$ and $2 R_\rmd$, respectively. The image reveals a general decline in metallicity with increasing radial distance, indicating a negative metallicity gradient. The right-hand panel displays the mass-weighted histogram of $Z$ for cold gas cells located between $r = R_\rmd$ and $r = 2 R_\rmd$, and vertically between $z = -H_\rmd$ and $z = H_\rmd$, normalized by the total mass of cold gas in the specified volume. The black dashed vertical line at $Z = 0.009$ marks the $99^{\rm th}$ percentile of this distribution and serves as the threshold metallicity, $Z_{\rm cut}$, used to distinguish between stream gas ($Z < Z_{\rm cut}$) and non-stream gas ($Z > Z_{\rm cut}$). This stream selection procedure is based on the assumption that the vast majority of cold gas located beyond $R_\rmd$ originates from recently accreted streams.}
    \label{fig:met_sel}
\end{figure*}

\begin{figure*}
    \centering
    \includegraphics[width=0.95\textwidth]{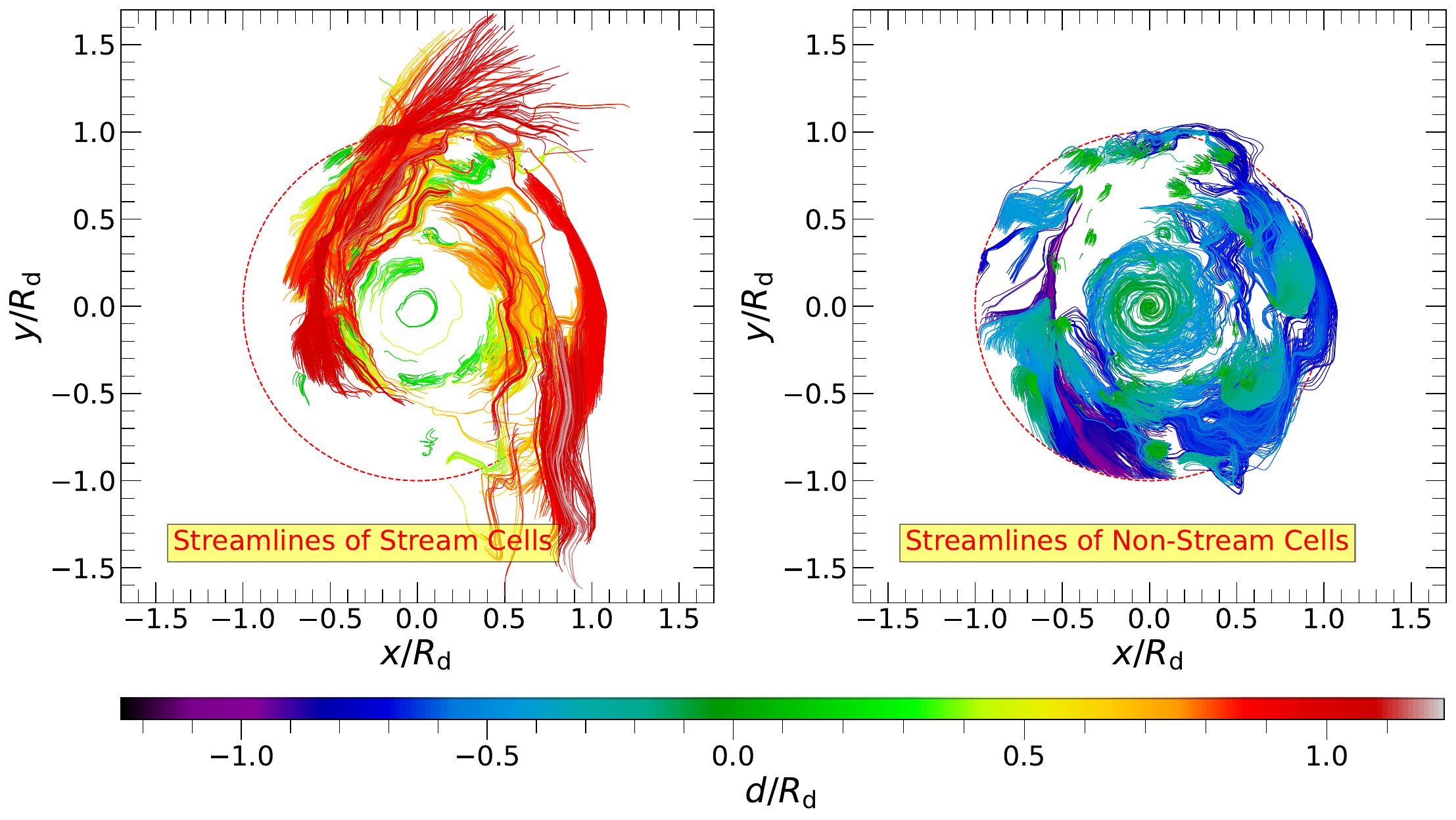}
    \caption{{\bf Streamline-Based Selection of Streams}: For the VELA 7 galaxy at a redshift $z=1.5$, the figure displays pathlines for a randomly selected subset of cold gas cells within the galactic disk, traced backward over an average disk dynamical timescale, defined as $t_{\rm trace} = R_\rmd / \langle V_{\rm rot} \rangle$, and projected onto the $x$–$y$ plane. These pathlines are shown separately for streams (left-hand panel) and non-streams (right-hand panel). To enhance visual clarity, only a portion of the cold gas cells within the disk are included. The pathlines are color-coded by $\pm d$, where $d$ represents the distance a gas cell has traveled during $t_{\rm trace}$, normalized by $R_\rmd$. Positive values of $d$ indicate inward motion (i.e., the 3D radius at $t - t_{\rm trace}$ is larger than at the present time), while negative values signify outward movement. Red dashed circles mark the radial boundary of the disk. A threshold value of $d_{\rm cut} = +0.1 R_\rmd$ is used to distinguish inflowing streams from non-streaming gas. Most stream-classified cells (left-hand panel) cluster into two spatially coherent inflowing structures, both showing large positive $d$ values and entering the disk either from the top or bottom-right regions of the projection. In contrast, cells labeled as non-streams (right-hand panel) either show minimal displacement over $t_{\rm trace}$ ($d \approx 0$) or exhibit outward motion with significantly negative $d$ values.}
    \label{fig:sl_sel}
\end{figure*}

\begin{figure*}
    \centering
    \includegraphics[width=0.95\textwidth]{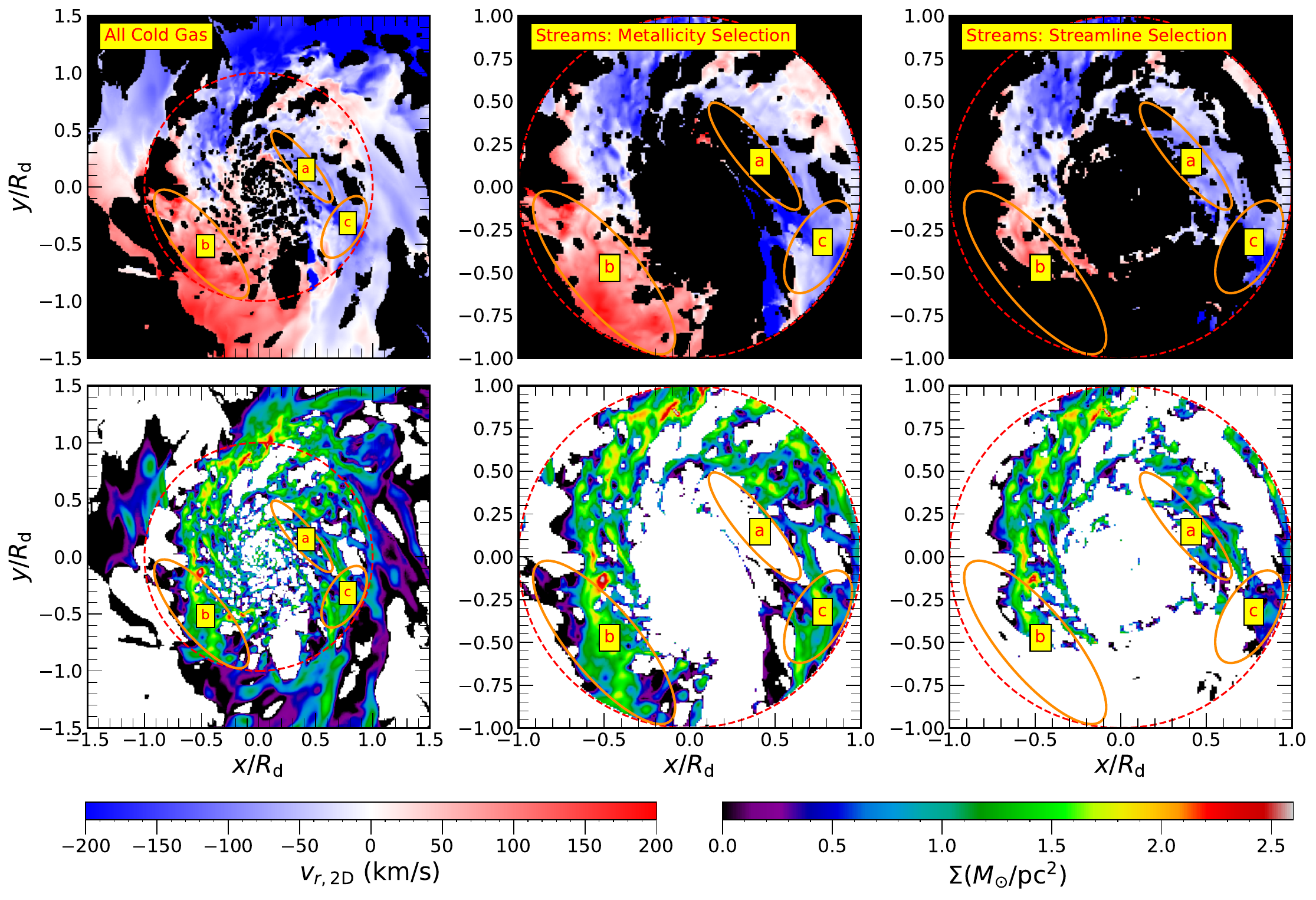}
    \caption{{\bf Stream Maps}: The left-, middle, and right-hand panels respectively present the face-on maps of radial velocity (top row) and surface density (bottom row) in VELA 7 at $z=1.5$, corresponding to all cold gas, streams identified via the metallicity-based method, and streams identified via the streamline-based method. To aid visual recognition of streams, the maps in the left-hand panels are shown over an expanded area of $3 R_\rmd \times 3 R_\rmd$. However, the actual classification of cold gas into streams and non-streams is confined to the galactic disk, defined as a cylinder with radius $R_\rmd$ and height $H_\rmd$, as shown in the middle and right-hand panels. Red dashed circles denote the radial extent of this disk. In the left-hand panels, two distinct inflowing cold gas groups are apparent (blue in the $v_{r, \rm 2D}$ map indicates inward motion), entering within $R_\rmd$ either from the top or the bottom-right and subsequently spiraling counter-clockwise toward the galactic center. The metallicity- and streamline-based classifications select somewhat different portions of these inflowing groups as streams. The areas enclosed by the orange ellipses (labeled a, b, and c) correspond to the main regions of disagreement between the two classification methods; these discrepancies are discussed further in \se{st_sig_ex}}
    \label{fig:image_st}
\end{figure*}

\subsection{Comparison with Simulations}
\label{sec:compare_to_theory}

In Fig.~\ref{fig:compare_th}, the predictions for the radial inflow velocity from the models discussed above are compared against the results from the VELA simulations. For each snapshot in our disk galaxy sample, $\langle V_r \rangle/\langle V_{\rm rot} \rangle$ is plotted against $\langle V_{\rm rot} \rangle/\langle \sigma_r \rangle$ with the blue, round points. The red squares and the vertical error bars denote the medians and $16^{\rm th}-84^{\rm th}$ percentiles, respectively, in $\langle V_{\rm rot} \rangle/\langle \sigma_r \rangle$ bins of varying widths, as indicated by the horizontal error bars, with the locations of the squares along the $\langle V_{\rm rot} \rangle/\langle \sigma_r \rangle$ - axis being at the bin-averages. The different curves show the inflow velocity predictions from the disk-instability-driven models of radial transport, as labeled, with $Q=0.68$ \citep[marginal instability condition for thick disks;][]{goldreich65} in the left-hand panel and $Q=1$ (marginal instability condition for thin disks) in the right-hand panel.

In the simulations, on average, $\langle V_r \rangle/\langle V_{\rm rot} \rangle$ decreases in magnitude as $\langle V_{\rm rot} \rangle/\langle \sigma_r \rangle$ increases. While the different models reflect this general trend, they typically produce higher levels of radial inflow than observed in the simulations. With $Q=0.68$ (left-hand panel), for $\langle V_{\rm rot} \rangle/\langle \sigma_r \rangle \lesssim 5$, all the models predict significantly higher radial velocities (in magnitude) than generally found in the VELA simulations. For $\langle V_{\rm rot} \rangle/\langle \sigma_r \rangle > 5$, however, the \citetalias{dekel20} and \citetalias{krumholz10} models agree with the simulations. But, there are very few galaxies with such high $\langle V_{\rm rot} \rangle/\langle \sigma_r \rangle$ (only about $20 \%$). With $Q=1$ (right-hand panel), the different model curves are closer to the simulated points, but still, for $\langle V_{\rm rot} \rangle/\langle \sigma_r \rangle \lesssim 5$, where most of the simulated points lie, only the \citetalias{dekel20} model is in good agreement with the simulations. However, in the case of $\langle V_{\rm rot} \rangle/\langle \sigma_r \rangle > 5$, all the models with $Q=1$ agree with the simulations.

As noted earlier, the \citetalias{dekel20} model specifically concerns the migration of thin rings, but the fraction of VELA disks with strong rings is small \citepalias[$\lesssim 30 \%$, see][]{dekel20}. The \citetalias{dekel09} and \citetalias{dekel13} models apply to the migration of clumps. However, we have verified that the radial mass flux in the VELA disks is dominated by the off-clump material. Therefore, the \citetalias{krumholz10} and \citetalias{krumholz18} models, which deal with global gas transport in viscous disks, may be the most relevant for us. From Fig.~\ref{fig:compare_th}, we can safely conclude that for most of the VELA disks, irrespective of whether $Q$ is assumed to be $0.68$ or $1$, both these models overpredict the magnitudes of the disk-averaged radial inflow velocities compared to that found in the simulations (see also Appendix~\ref{sec:A3}, which shows that the same is true for the locally averaged inflow at different radii as well). We also emphasize that all five models predict radial inflows, whereas occasionally (in about $18 \%$ cases), disk-averaged outflows can be seen in the VELA simulations. 


\section{Streams}
\label{sec:st}

As seen in Fig.~\ref{fig:ex}, the VELA galaxies are fed by non-axisymmetric streams of cold gas, which can penetrate well within the disk. As such, much of the radial mass flux and velocity associated with the VELA disks could well be dominated by the motion of recently accreted streams rather than being generated by disk instability. To ascertain if this is indeed the case, we first compare the radial mass flux at the boundary of VELA disks (in a cylindrical shell of radius $r=R_{\rm d}$, width $\Delta r=0.1 R_{\rm d}$, and height, $H_{\rm d}$) to the cold gas accretion rate at the disk boundary predicted by a simple ``bathtub" model \citep[e.g.,][]{bouche10,dekel14} in steady state, where the star formation rate in the disk is balanced by the accretion rate, ignoring any mass inflows through the disk onto the central bulge or mass outflows (\se{bathtub}). We then discuss how streams are identified in the VELA disks (\se{selection}) and quantify the radial transport of streams and off-stream cold gas, labeled as non-streams, separately. We highlight a particular example in \se{st_sig_ex} and discuss the overall statistics from the VELA disk galaxy sample in \se{st_sig_stats}.

\subsection{Radial Mass Flux vs Accretion Rate from Bathtub Model}
\label{sec:bathtub}

Assuming mass conservation, the total cold gas mass, $M_{\rm d}$, within a disk varies according to
\begin{equation}
    \dot{M}_{\rm d}=\dot{M}_{\rm ac}-\dot{M}_{\rm sf} \,,
    \label{eq:bath}
\end{equation}
where $\dot{M}_{\rm ac}$ is the accretion rate of cold gas at the disk edge and $\dot{M}_{\rm sf}$ is the star formation rate, crudely approximated as 
\begin{equation}
\dot{M}_{\rm sf}=M_{\rm d}/\tau_{\rm sf} \,,
\label{eq:sf}
\end{equation}
where $\tau_{\rm sf}=t_{\rm ff}/\epsilon_{\rm sf}$. Here, $t_{\rm ff}$ is the free-fall time, and $\epsilon_{\rm sf}$ is the star formation efficiency. Note that while deriving Equation~\ref{eq:bath}, we have neglected all sink terms other than star formation, such as disk-instability driven mass inflows through the disk onto the central bulge and outflows driven by stellar feedback.

If $\dot{M}_{\rm ac}$ and $\tau_{\rm sf}^{-1}$ are slowly varying with timescales that are longer than $\tau_{\rm sf}$, Equations~\ref{eq:bath} and \ref{eq:sf} can be easily solved to give
\begin{equation}
    M_{\rm d}=\dot{M}_{\rm ac}\tau_{\rm sf}(1-e^{-t/\tau_{\rm sf}}) \,, \ \ \ \dot{M}_{\rm d}=\dot{M}_{\rm ac} e^{-t/\tau_{\rm sf}} \,.
\end{equation}
After a transition period of order $\tau_{\rm sf}$, the above equation relaxes to a steady-state solution given by
\begin{equation}
    M_{\rm d} \simeq \dot{M}_{\rm ac}\tau_{\rm sf} \,, \ \ \ \dot{M}_{\rm d} \simeq 0 \,.
\end{equation}
Therefore, $\dot{M}_{\rm ac}/M_{\rm d} \simeq \tau_{\rm sf}^{-1} = \epsilon_{\rm sf} t_{\rm ff}^{-1}$.

Following \citetalias{dekel13}, we estimate $t_{\rm ff}$ as $t_{\rm ff}=f_{\rm sf} t_{\rm d}$, where $f_{\rm sf} \sim 0.5$, assuming that star formation occurs in clumps where the overdensity is a factor of few with respect to the average disk density and 
\begin{equation}
    t_{\rm d}=R_{\rm d}/V \simeq \upsilon^{-1} \lambda R_{\rm v}/V_{\rm v} \simeq 0.14  \upsilon^{-1} \lambda t
\end{equation}
is the disk dynamical time. Here, $\lambda$ is the halo spin parameter, relating the disk radius, $R_\rmd$, to the halo virial radius, $R_{\rm v}$, $\upsilon$ is the ratio between the disk rotation velocity, $V$, and the halo virial velocity, $V_{\rm v}$, and using the redshift dependent relations for $R_{\rm v}$ and $V_{\rm v}$, $R_{\rm v}/V_{\rm v}=0.14 t$, where $t$ is the cosmological time. Adopting $\lambda \sim 0.05$, $\upsilon \sim 1$, and $\epsilon_{\rm sf} \sim 0.02$, we get
\begin{equation}
    \dot{M}_{\rm ac}/M_{\rm d} \simeq \frac{1}{0.175t} = \frac{a^{-3/2}}{0.175 t_1} \,,
    \label{eq:bathtub}
\end{equation}
where we have used the relation $1/a=(t/t_1)^{-2/3}$, with $t_1=17.5\ {\rm Gyr}$ \citepalias[see][]{dekel13}. For comparison, the cosmological specific accretion rate of baryons at the halo boundary is given by
\begin{equation}
    \frac{\dot{M}_{\rm b}}{M_{\rm b}} \simeq s a^{-5/2} \,,
    \label{eq:sar}
\end{equation}
with $s=0.03\ {\rm Gyr}^{-1}$, which was also derived in \citetalias[][]{dekel13} and shown to be in good agreement with earlier VELA-like simulations.

In Fig.~\ref{fig:bathtub}, the cold gas radial mass flux at $R_\rmd$, $F_{r=R_\rmd}$, divided by $M_\rmd$ is shown as a function of the scale factor $a$ for our disk galaxy sample, with the solid black line and the envelope indicating the median and the $16^{\rm th}-84^{\rm th}$ percentile variation over $a$ bins of width $0.07$. The brown, dot-dashed line plots $\dot{M}_{\rm ac}/M_{\rm d}$ as a function of $a$ according to Equation~\ref{eq:bathtub}. For comparison, the blue, dotted line also indicates $\dot{M}_{\rm b}/M_{\rm b}$ as a function of $a$ according to Equation~\ref{eq:sar}. The good agreement between the median trend in $F_{r=R_\rmd}/M_{\rm d}$ and $\dot{M}_{\rm ac}/M_{\rm d}$ from the bathtub model, shows that on average, the cold gas radial mass flux at the edge of VELA disks is consistent with that expected from accretion onto the disk. In the following subsections, we show that in VELA disks, the accreted cold gas streams can penetrate well within the disk, such that the radial mass flux in the outer disks ($r \gtrsim 0.5 R_{\rmd}$) is largely dominated by stream motions.

\subsection{Stream Selection}
\label{sec:selection}

The identification of gas in incoming streams is not a trivial task given the Eulerian nature of the VELA simulations and the absence of gas particles that trace the flows. We, therefore, appeal to two different crude procedures for selecting grid cells that belong to recently incoming streams, one using gas metallicities and the other using streamlines, as follows. 

\subsubsection{Metallicity-Based Selection}
\label{sec:met_sel}

Streams, due to their recent accretion onto the galaxy, are likely to have relatively low metallicities. Therefore, in a crude way, we can use a threshold metallicity, $Z_{\rm cut}$, to separate streams from the rest of the cold gas. To determine this metallicity threshold for a particular snapshot, we first find the mass-weighted cold gas metallicity distribution outside the disk radius, between $r=R_\rmd$ to $2 R_\rmd$ and $z=-H_\rmd$ to $H_\rmd$, and then select the $99^{\rm th}$ percentile of this distribution as $Z_{\rm cut}$. Our results are robust against the particular choice for this threshold, as adopting $Z_{\rm cut}$ to be the $90^{\rm th}$ or $95^{\rm th}$ percentile instead of the $99^{\rm th}$ has no significant impact (see Appendix~\ref{sec:A2}). Cold gas with $Z<Z_{\rm cut}$ at any $r$ is classified as streams and the rest as non-streams. The inherent assumption made in this selection procedure is that almost all the cold gas beyond $R_\rmd$ is in the form of recently accreted streams. Therefore, the maximum metallicity of this gas is the maximum metallicity of streams and can be used as a threshold to distinguish between streams and non-streams within the disk. Note that here we ignore the possibility that the gas beyond $R_\rmd$ may also have a metal-enriched, outflowing cold component ejected by stellar winds and supernovae \citep[e.g.][]{veilleux20}, which can potentially bias this selection. However, the inflowing cold streams may themselves have higher metallicities because of the entrainment of initially hot, recycled, enriched gas that has been previously ejected from the galaxy \citep[][]{strawn21, aung24}.

To give an example, in Fig.~\ref{fig:met_sel}, the left-hand panel shows the 2D metallicity map, $Z_{\rm 2D}$, for VELA 7 at $z=1.5$ in a face-on projection of dimensions $4 R_\rmd \times 4 R_\rmd$, centered on the galactic center, obtained after mass-weighted averaging the metallicites ($Z$) of the cold gas cells along the $z$-axis between $-H_\rmd$ to $H_\rmd$. The red and cyan, dashed circles have radii equal to $R_\rmd$ and $2 R_\rmd$, respectively. The image shows an overall negative metallicity gradient from small to large $r$, i.e., metallicity decreases as the cylindrical distance from the galactic center increases. This is consistent with the assumption that the cold gas outside the disk ($r \gtrsim R_\rmd$) likely consists almost entirely of low-metallicity streams, whereas that within the disk ($r \lesssim R_\rmd$) is a mix of both streams and non-streams. In the right-hand panel, the mass-weighted histogram of $Z$ for all cold gas cells between $r=R_\rmd$ to $2 R_\rmd$ and $z=-H_\rmd$ to $H_\rmd$ is shown, normalized by the total cold gas mass inside this region. The black, dashed vertical line at $Z=0.009$ indicates $Z_{\rm cut}$, the $99^{\rm th}$ percentile of this distribution, and the threshold metallicity that separates streams from non-streams.

\subsubsection{Streamline-Based Selection}
\label{sec:sl_sel}

Alternatively, we try to identify streams by tracing the path of each cold gas cell back in time. For simplicity, we assume that the gas velocity field remains stationary during the time over which we trace back the flow. Therefore, pathlines are the same as instantaneous streamlines, and cold gas cells in a given snapshot can be traced using the gas velocity data of that snapshot only. The tracing time, $t_{\rm trace}$, is adopted to be the average disk dynamical time, $t_{\rm trace} = R_\rmd / \langle V_{\rm rot} \rangle$. Starting from its current 3D position, $\vec{r_{t}}$, at time, $t$, a cold gas cell in a given snapshot is traced back in time along its pathline (i.e., the streamline that originates from this position) to yield its initial 3D position, $\vec{r_{i}}$, at time $t-t_{\rm trace}$. As streams are recently accreted material, the cells which move inwards over $t_{\rm trace}$, i.e., $|\vec{r_{i}}|>|\vec{r_{t}}|$, and by sufficiently large distances, i.e., $d=|\vec{r_{t}}-\vec{r_{i}}|>d_{\rm cut}$, are classified as streams. The rest are classified as non-streams. By trial and error, we choose $d_{\rm cut} = 0.1 R_\rmd$, such that most cold gas cells that visually appear to be a part of streams get correctly identified by this procedure. Changing $d_{\rm cut}$ by a factor of two up or down has no significant impact on our results (see Appendix~\ref{sec:A2}). We note that the streamline-based selection is biased towards detecting only inflowing material as streams, while the metallicity-based selection doesn't have this bias.

As an example, for VELA 7 at $z=1.5$, the pathlines of a random selection of cold gas cells within the disk, traced back in time over the interval $t_{\rm trace}$ and projected onto the $x$-$y$ plane, are shown in Fig.~\ref{fig:sl_sel}, separated into streams (left-hand panel) and non-streams (right-hand panel). The pathlines of all cold gas cells within the disk are not shown for better clarity. The lines are color-coded by $\pm d$, in units of $R_\rmd$, with positive values indicating the cases where $|\vec{r_{i}}|>|\vec{r_{t}}|$ and negative values showing the opposite. The red circles denote the radial extent of the disk. We note that most of the pathlines in the left-hand panel can be divided into two distinct groups, both having large positive $d$ values, indicating the inflow of cold gas either from the top or bottom-right of the image, which then spiral onto the galaxy counter-clockwise. The two groups also broadly coincide in space with the regions of the galaxy that can be visually identified as two distinct streams in the radial velocity and surface density maps (see the top left- and right-hand panels of Fig.~\ref{fig:ex}). On the other hand, the pathlines in the right-hand panel either have $d \approx 0$ (greenish lines), indicating little or no displacement over $t_{\rm trace}$, or have large negative $d$ values, indicating outflowing cold gas.

\subsection{Stream and Non-Stream Signal: An Example}
\label{sec:st_sig_ex}

\begin{figure*}
    \centering
    \includegraphics[width=0.95\textwidth]{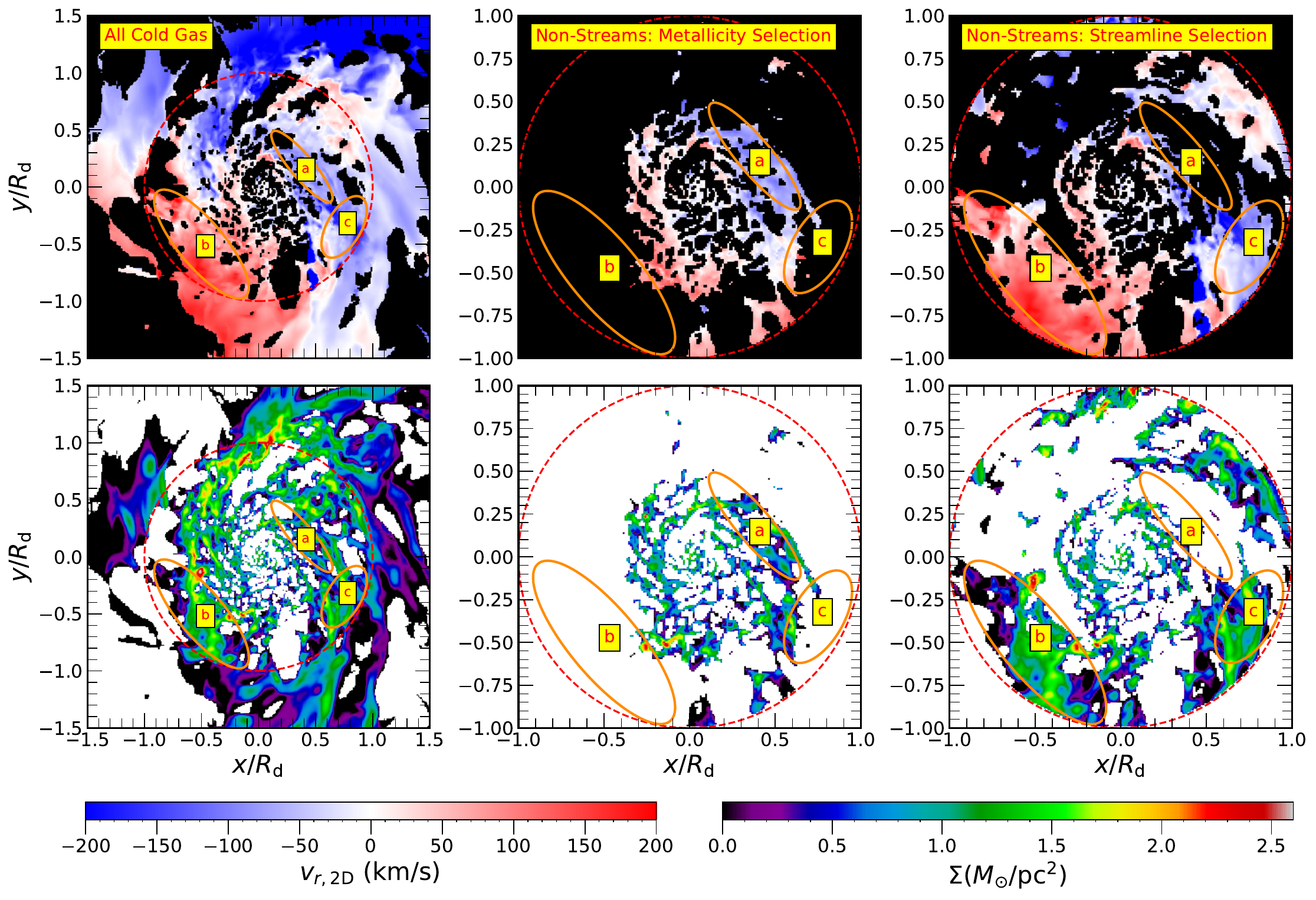}
    \caption{{\bf Non-Stream Maps}: Same as in Fig.~\ref{fig:image_st}, except the middle and right-hand panels now show the face-on radial velocity (top row) and surface density (bottom row) maps of the off-stream material obtained with the metallicity- and streamline-based selections, respectively. As seen in Figure~\ref{fig:image_st}, the metallicity- and streamline-based selections identify somewhat different regions of the cold gas as streams, automatically resulting in different non-stream identifications. The regions marked with the orange ellipses (a, b, and c) are the primary sources of disagreement between the two classification methods as discussed in \se{st_sig_ex}.}
    \label{fig:image_nst}
\end{figure*}

\begin{figure*}
    \centering
    \includegraphics[width=0.95\textwidth]{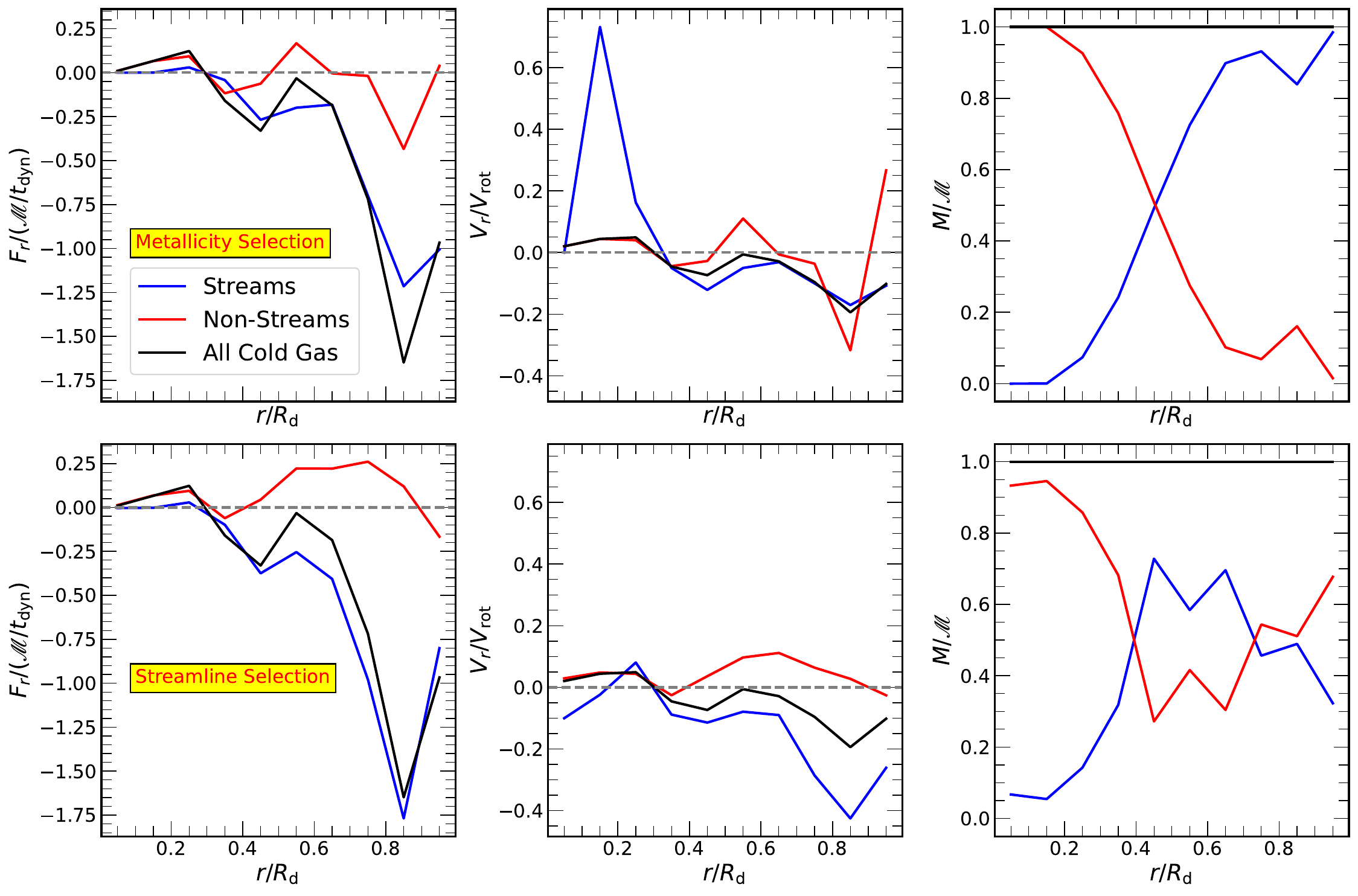}
    \caption{{\bf Stream and Non-Stream Radial Profiles}: The left-, middle, and right-hand panels respectively show the radial mass flux ($F_r$, in units of $\mathscr{M} / t_{\rm dyn}$), average radial velocity ($V_r$, in units of $V_{\rm rot}$), and mass ($M$, in units of $\mathscr{M}$), as functions of radius $r$, normalized by $R_\rmd$, for all cold gas (black), streams (blue), and non-streams (red) in VELA 7 at $z=1.5$. The top row presents results from the metallicity-based classification, and the bottom row corresponds to the streamline-based classification. In both cases, the total radial mass flux of cold gas is dominated by streams at large radii (beyond $\sim \! 0.7\ R_\rmd$) and by non-streams at small radii (within $\sim \! 0.3\ R_\rmd$). In the metallicity-based selection, the dominance of stream flux at large $r$ primarily reflects the higher mass fraction of streams in that region. In contrast, the streamline-based method shows roughly equal masses for streams and non-streams at large $r$; however, the non-stream mass flux remains low. This is because the inflowing and outflowing contributions from non-stream gas nearly cancel out (see Fig.~\ref{fig:image_nst}), leaving streams as the main net contributor to the total radial mass flux.}
    \label{fig:profiles}
\end{figure*}

\begin{figure*}
    \centering
    \includegraphics[width=0.96\textwidth]{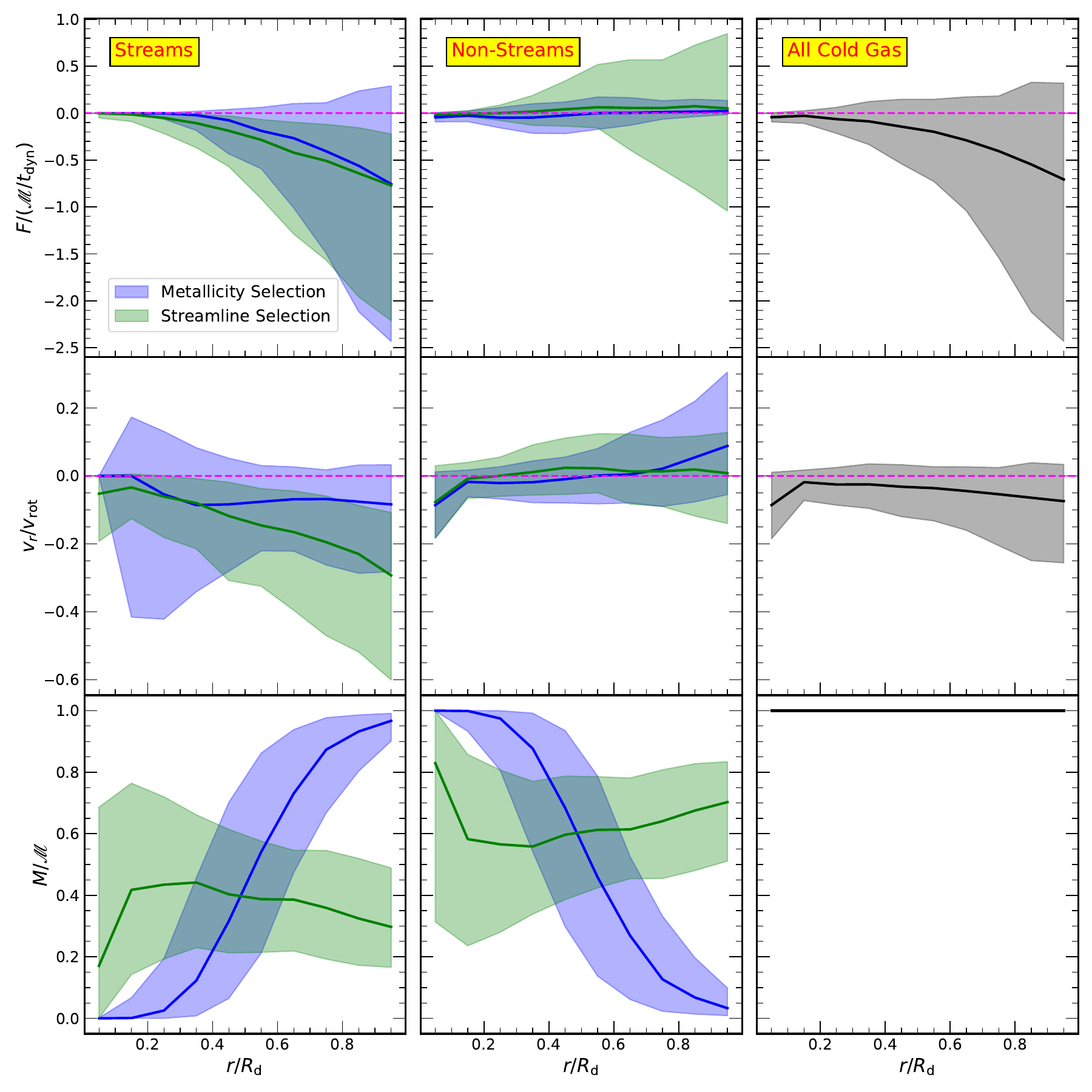}
    \caption{{\bf Stream and Non-Stream Statistics}: The top, middle, and bottom rows display, respectively, the radial mass flux ($F_r$, in units of $\mathscr{M}/t_{\rm dyn}$), average radial velocity ($V_r$, in units of $V_{\rm rot}$), and mass ($M$, in units of $\mathscr{M}$), as functions of radius $r$, normalized by $R_\rmd$, for streams (left panels), non-streams (middle panels), and all cold gas (right panels). Solid lines represent the median values across the disk galaxy sample, while shaded regions indicate the $16^{\rm th}$–$84^{\rm th}$ percentile ranges. The results from the metallicity-based selection are shown in blue, and those from the streamline-based selection are shown in green. In both classifications, the median cold gas mass flux at large radii is dominated by streams — beyond $r \gtrsim 0.5 R_\rmd$ in the metallicity-based selection and beyond $r \gtrsim 0.3 R_\rmd$ in the streamline-based selection. At smaller radii, the median stream flux approaches zero and falls below that of the non-stream component, resulting in a total cold gas flux that, while small, is primarily contributed by non-streams. The median radial velocity of streams remains predominantly negative across all radii, and its magnitude tends to exceed that of the non-streams, particularly at large $r$. Thus, on average, the radial transport of cold gas in the outer regions of the VELA disk galaxy sample is driven primarily by the infall of recently accreted streams.}
    \label{fig:stats}
\end{figure*}

\begin{figure*}
    \centering
    \includegraphics[width=0.96\textwidth]{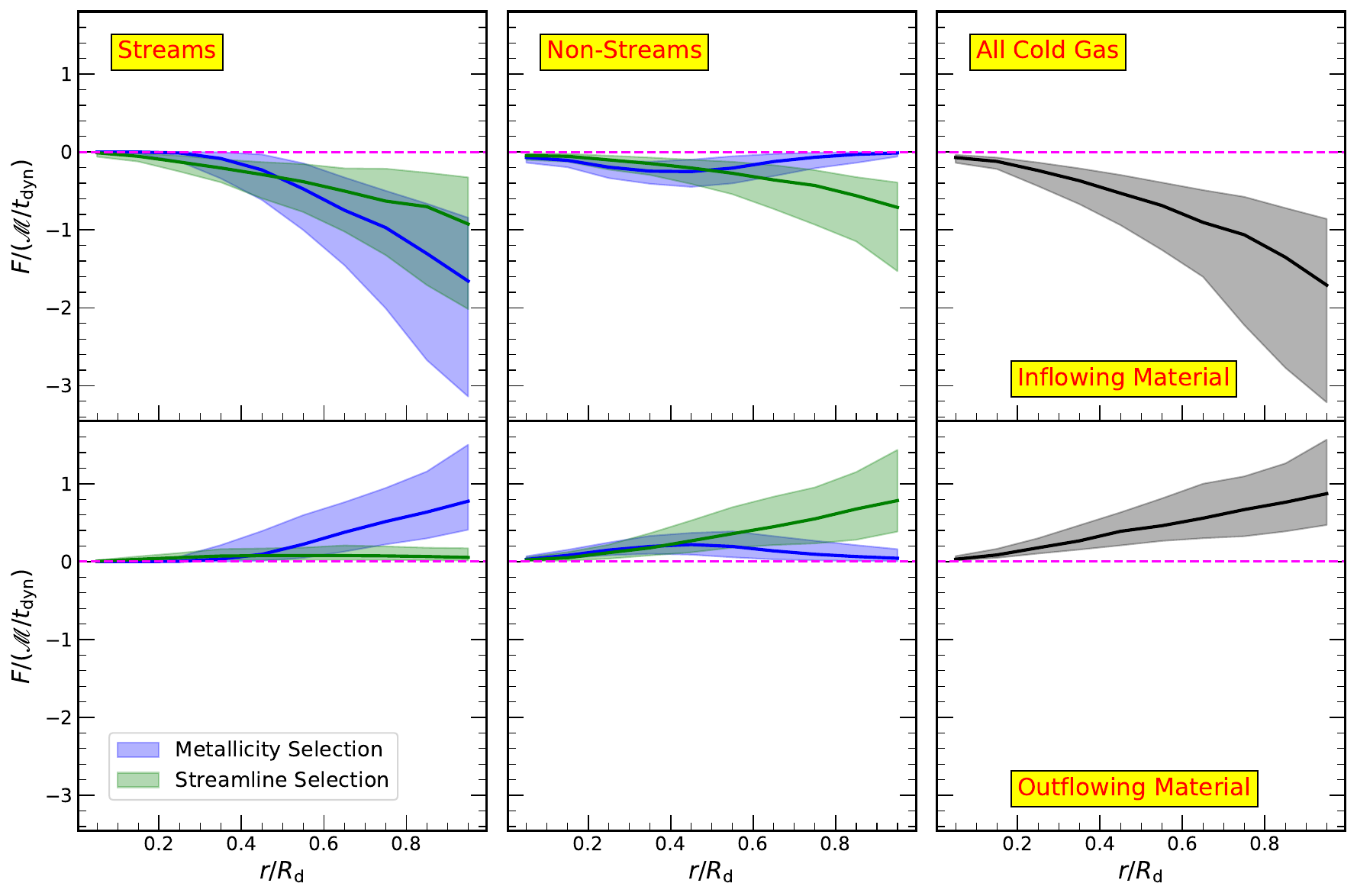}
    \caption{{\bf Positive vs Negative Flux}: The top and bottom rows illustrate the radial mass flux ($F_r$, in units of $\mathscr{M}/t_{\rm dyn}$), from inflowing and outflowing regions, respectively, for streams (left-hand panels), non-streams (middle panels), and the total cold gas (right-hand panels), plotted as a function of radius $r$, normalized by $R_\rmd$. Solid lines represent the median values across the disk galaxy sample, and shaded regions correspond to the $16^{\rm th}$–$84^{\rm th}$ percentile ranges. Results from the metallicity- and streamline-based classifications are shown in blue and green, respectively. In the streamline-based selection, inflowing and outflowing fluxes from non-stream gas are nearly equal in magnitude, comparable also to the inflowing stream flux, and largely cancel each other out. This leads to a total non-stream flux that is close to zero at the median level, though it exhibits significant scatter. A similar cancellation is also observed in the metallicity-based selection, where inflowing and outflowing non-stream fluxes are also comparable but individually much smaller than the stream flux, except at small radii ($r \lesssim 0.5\ R_\rmd$), where the stream flux diminishes. As a result, the total non-stream flux in the metallicity-based selection also approaches zero, but with minimal scatter.}  
    \label{fig:stats_pos_neg}
\end{figure*}

In Figs.~\ref{fig:image_st} and \ref{fig:image_nst}, we compare the face-on radial velocity (top rows) and surface density (bottom rows) maps of streams and non-streams, respectively, obtained with the metallicity- (middle panels) and streamline-based (right-hand panels) selections for VELA 7 at $z=1.5$. These maps are generated following the procedure outlined in \se{ex}, except that instead of all cold gas cells, only cells identified as streams or non-streams are used, as relevant. To check the results of the two selections against what visually appears as streams, the face-on radial velocity and surface density maps of all cold gas in a slightly larger region of dimensions $3 R_\rmd \times 3 R_\rmd$, as shown in the top right- and left-hand panels of Fig.~\ref{fig:ex}, are re-plotted here in the top and bottom left-hand panels, respectively. Note that the actual classification of cold gas cells into streams and non-streams is only done within the disk (i.e., within a cylinder of radius $R_\rmd$ and height $H_\rmd$), as shown in the middle and right-hand panels. The red, dashed circles indicate the radial extent of the disk. 

As visually identified earlier in Fig~\ref{fig:ex}, there are two distinct groups of incoming cold gas (blue indicates inflows in the $v_{r, \rm 2D}$ map), one entering within $R_\rmd$ from the top (Group I) and the other from the bottom-right (Group II) of the maps, which then spiral onto the galaxy counter-clockwise. Because of this spiraling motion, the bulk of the red, outflowing region in the bottom-left portion of the maps may be interpreted as a part of Group I, en route to the apocenter after having crossed the pericenter. The metallicity-based selection identifies most of the cold gas in these two groups as streams (middle panels of Fig.~\ref{fig:image_st}), except for a portion of Group II (region a), which doesn't get selected because of its somewhat higher metallicity (see the left-hand panel of Fig.~\ref{fig:met_sel}) and remains as a part of non-streams (middle panels of Fig.~\ref{fig:image_nst}). This material, however, gets identified as streams in the streamline-based selection (right-hand panels of Fig.~\ref{fig:image_st}), as the cold gas cells here have inflowing pathlines with large positive $d$ values (see the left-hand panel of Fig.~\ref{fig:sl_sel}). But contrary to the metallicity-based selection, most of the outflowing cold gas in Group I (region b) does not get recognized as streams in the streamline-based selection, as their pathlines are both currently and on-average outflowing over the tracing time, with negative $d$ values (see the right-hand panel of Fig.~\ref{fig:sl_sel}). Similarly, the cold gas in region c, which is currently inflowing, is not identified as streams in the streamline-based selection, as it is, on average, outflowing over the tracing time. While geometrically coincident with Group II, this material is likely to be kinematically associated with Group I as it falls back into the galaxy after its apocentric passage and is not a part of the streamline group that enters within $R_\rmd$ from the bottom-right (of the maps). As such, the cold gas here has been likely circulating in the galaxy for quite some time as opposed to being recently accreted and, therefore, perhaps rightfully, should not be classified as streams. 

The statement above notwithstanding, we note that both of our stream selection procedures are rather crude, and it is difficult to ascertain their comparative accuracy. Therefore, the somewhat different stream and non-stream classifications of cold gas resulting from the metallicity- and streamline-based selections are accepted at face value. In \se{st_sig_stats}, we will find that, on average, the mass in streams in the metallicity-based selection is more than that in the streamline-based selection. Therefore, the former can be regarded as a liberal stream selection procedure, while the latter is more conservative. 

In Fig.~\ref{fig:profiles}, for VELA 7 at $z=1.5$, the left-, middle, and right-hand panels, respectively, show the radial mass flux, $F_r$ (in units of $\mathscr{M}/t_{\rm dyn}$), average radial velocity, $V_r$ (in units of $V_{\rm rot}$), and mass, $M$ (in units of $\mathscr{M}$), as functions of $r$ (in units of $R_\rmd$) for all cold gas (black), streams (blue), and non-streams (red). The results from the metallicity-based selection are shown in the top row, while those from the streamline-based selection are plotted in the bottom row. Note that the cold gas $V_r$ and $F_r$ are re-plotted from the bottom left- and right-hand panels of Fig.~\ref{fig:ex}, respectively, for convenient comparison with the stream and non-stream signals.

For the metallicity-based selection (top row), the mass in streams (non-streams) increases (decreases) with an increase in $r$, attaining values close to $\mathscr{M}$ (zero) near $r=R_\rmd$. Consequently, the cold gas radial mass flux, which is a sum of the stream and non-stream flux, is also dominated by the streams at large $r$, outside $\sim \! 0.7\ R_\rmd$. Between $\sim \! 0.3\ R_\rmd$ to $\sim \! 0.7\ R_\rmd$, the stream and non-stream fluxes are comparable in magnitude, with the non-stream flux oscillating between positive and negative values, while the stream flux is always negative. Inside $\sim \! 0.3\ R_\rmd$, the stream flux tends to zero, and the cold gas flux, albeit very small, is dominated by the non-streams and directed outwards. The stream $V_r$ is mostly negative (except within $\sim \! 0.3\ R_\rmd$), reaching values close to $-0.2\ V_{\rm rot}$ near $r=R_\rmd$. The non-stream $V_r$ oscillates between positive and negative values. The cold gas $V_r$, which is a mass-weighted average of the stream and non-stream $V_r$, is almost the same as the stream $V_r$ outside $\sim \! 0.7\ R_\rmd$ and as the non-stream $V_r$ inside $\sim \! 0.3\ R_\rmd$, owing to the mass dominance of the streams and non-streams in these two regions, respectively.

For the streamline-based selection (bottom row), the mass in streams (non-streams) increases (decreases) with $r$ only up to about $\sim \! 0.4\ R_\rmd$, after which it roughly stays the same up to $\sim \! 0.7\ R_\rmd$, and then decreases (increases). The cold gas radial mass flux outside $\sim \! 0.7\ R_\rmd$ is dominated by the streams, as the non-stream flux is much smaller in magnitude. However, note that this is because of outflowing and inflowing non-stream fluxes of comparable magnitude almost canceling each other (see right-hand panels of Fig.~\ref{fig:image_nst}; contributions to the total non-stream flux from regions b and c almost cancel each other) and not because of small non-stream mass as in the metallicity-based selection. Similar to the metallicity-based selection, between $\sim \! 0.4\ R_\rmd$ to $\sim \! 0.7\ R_\rmd$, the stream and non-stream fluxes are comparable in magnitude, with the non-stream flux being mostly positive and the stream flux being negative. Between $\sim \! 0.3\ R_\rmd$ to $\sim \! 0.4\ R_\rmd$, however, the stream flux again dominates, but inside $\sim \! 0.3\ R_\rmd$, the stream flux tends to zero, and the cold gas flux, albeit very small, is dominated by the non-streams as in the metallicity-based selection. The stream $V_r$ is mostly negative, reaching values close to $-0.4\ V_{\rm rot}$ near $r=R_\rmd$, while the non-stream $V_r$ is mostly positive and smaller in magnitude than the stream $V_r$. As the stream and non-stream masses are comparable outside $\sim \! 0.4\ R_\rmd$, the cold gas $V_r$, being a mass-weighted average of the stream and non-stream $V_r$, is somewhere in between (and negative). Inside $\sim \! 0.4\ R_\rmd$, the non-stream mass dominates, and the cold gas $V_r$ is close to the non-stream $V_r$.

\subsection{Stream and Non-Stream Signal: Statistics}
\label{sec:st_sig_stats}

After having discussed a specific example, we now focus on the statistical findings from the VELA disk galaxy sample. In Fig.~\ref{fig:stats}, the top, middle, and bottom rows, respectively, show $F_r$ (in units of $\mathscr{M}/t_{\rm dyn}$), $V_r$ (in units of $V_{\rm rot}$), and $M$ (in units of $\mathscr{M}$) as functions of $r$, in units of $R_\rmd$, for streams (left-hand panels), non-streams (middle panels), and all cold gas (right-hand panels), with the solid lines indicating the medians over the disk galaxy sample and the envelopes representing the $16^{\rm th}-84^{\rm th}$ percentiles. The results for the metallicity-based selection are shown in blue, while those for the streamline-based selection are plotted in green. Note that the $F_r$ and $V_r$ statistics for all cold gas (top and middle right-hand panels, respectively) are the same as those shown in Fig.~\ref{fig:stats_r}, re-plotted here for convenient comparison with the stream and non-stream statistics.

Most panels show a large scatter in the plotted quantities. Nonetheless, the median trends are illuminating. For the metallicity-based selection, the mass in streams (non-streams) increases (decreases) with increasing $r$, with the median mass attaining a value close to $\mathscr{M}$ (zero) near $r=R_\rmd$. In the streamline-based selection, the scatter is larger. However, the median stream (non-stream) mass increases (decreases) with increasing $r$, attains a maximum (minimum) of $\sim \! 0.4\ (0.6)\ \mathscr{M}$ around $\sim \! 0.4\ R_d$, and then slightly decreases (increases) to $\sim \! 0.3\ (0.7)\ \mathscr{M}$ near $r=R_\rmd$. Note that in the metallicity-based selection, the mass in streams (non-streams) dominates at large (small) $r$, with the median stream mass being greater than $0.5 \mathscr{M}$ for $r \gtrsim 0.5\ R_\rmd$. In the streamline-based selection, however, the mass in non-streams is greater than in streams (at least at the median level) at all $r$. Nevertheless, in both selections, the median non-stream flux is close to zero, and the median cold gas flux at large $r$ ($r \gtrsim 0.5\ R_\rmd$ in the metallicity-based selection and $r \gtrsim 0.3\ R_\rmd$ in the streamline-based selection) is dominated by the streams. At smaller $r$, the median stream flux tends to zero and is smaller than the median non-stream flux, such that the median cold gas flux, although very small, is dominated by the non-streams. The median stream $V_r$ is almost always negative, with typically larger magnitudes in the streamline-based selection than in the metallicity-based selection (the same is also true for the stream flux). The median non-stream $V_r$ is typically close to zero and smaller in magnitude than the median stream $V_r$, especially at large $r$. The median cold gas $V_r$ is always negative (also discussed in \se{stats}) but generally smaller in magnitude than the median stream $V_r$. This is expected as the cold gas $V_r$ is a mass-weighted average of the stream and non-stream $V_r$. From the median trends in $F_r$ and $V_r$, we conclude that, on average, the radial transport of cold gas in the VELA disk galaxy sample at large $r$ is dominated by the inflowing motion of recently accreted streams. These trends are also robust against the details of the feedback physics employed in the simulations, which we have verified by comparing these results with a newer generation of the VELA simulations \citep[][]{ceverino23} that uses the same set of initial conditions but has stronger stellar feedback (see Appendix~\ref{sec:A1}).

In Fig.~\ref{fig:stats}, we noted that the cold gas radial mass flux is dominated by the streams at large $r$ in both selections. For the metallicity-based selection, this can be understood from the fact that at large $r$, the streams dominate the total cold gas mass (see bottom row of Fig.~\ref{fig:stats}). However, in the streamline-based selection,  there is more mass in non-streams than in streams at all $r$ (at least at the median level). Despite this, the non-stream flux is close to zero. To investigate the reason behind this, in Fig.~\ref{fig:stats_pos_neg}, we separately show the radial mass flux (in units of $\mathscr{M}/t_{\rm dyn}$) from the inflowing (top row) and outflowing (bottom row) regions of the streams (left-hand panels), non-streams (middle panels), and all cold gas (right-hand panels) as a function of $r$ (in units of $R_\rmd$), with the solid lines indicating the medians over the disk galaxy sample and the envelopes representing the $16^{\rm th}-84^{\rm th}$ percentiles. The results for the metallicity-based selection are shown in blue, while those for the streamline-based selection are plotted in green. 

From the middle panels, we note that the inflowing and outflowing non-stream fluxes in the streamline-based selection are not negligible (see also \se{st_sig_ex}). In fact, they are similar in magnitude to the inflowing stream flux in the same selection. However, due to their comparable magnitude, they almost cancel each other, yielding a total non-stream flux that is close to zero at the median level, albeit with a large scatter (top-middle panel of Fig.~\ref{fig:stats}). The outflowing stream flux in this selection is negligible, as the selection procedure does not generally identify outflowing regions in the galaxy (which are also likely to be outflowing for the past dynamical time) as streams (see \se{st_sig_ex}). The inflowing and outflowing non-stream fluxes in the metallicity-based selection are also of comparable magnitude and can cancel each other. However, even by themselves, they are negligible compared to the stream flux in the same selection, except at small $r$ (inside $\sim \! 0.5\ R_\rmd$), where the stream flux is also small. This is because of the increasingly smaller mass in non-streams than in streams with increasing $r$. As such, the non-stream flux in this selection is also close to zero, with very little scatter (top-middle panel of Fig.~\ref{fig:stats}). The outflowing stream flux in the metallicity-based selection is not negligible, as in this selection, outflowing regions may get identified as streams based on their metallicity. However, we note that the outflowing stream flux is typically smaller in magnitude than the inflowing stream flux such that the total stream flux (top-left panel of Fig.~\ref{fig:stats}) is mostly negative. We also note that the non-stream inflowing and outflowing fluxes could just be non-circular motions that average out to zero. As we define inflows and outflows by the sign of the radial velocity, non-circular motions would be considered as flows in our analysis, and we can’t distinguish one from the other.

The marked difference in the stream mass in the two different selections, as seen in the bottom panel of Figure~\ref{fig:stats}, warrants some discussion. It is possible that the streamline-based selection is more reliable in the inner disk, due to increasing metallicities towards the galaxy centers, which may make it difficult to distinguish between streams and non-streams in the metallicity-based selection (see also Section~\ref{sec:st_sig_ex} and Figs.~\ref{fig:image_st} and \ref{fig:image_nst}). But even the streamline-based selection shows declining stream mass inside $\sim 0.2\ R_{\rmd}$, albeit not as stark as the decline in the metallicity-based selection. Over a few dynamical times, the streams are likely to phase-mix with the rest of the disk, and may become indistinguishable from the rest of the disk. However, more robust stream identification using tracer particles is required to understand the survival of streams within disks both temporally and spatially.

\begin{figure*}
    \centering
    \includegraphics[width=0.9\textwidth]{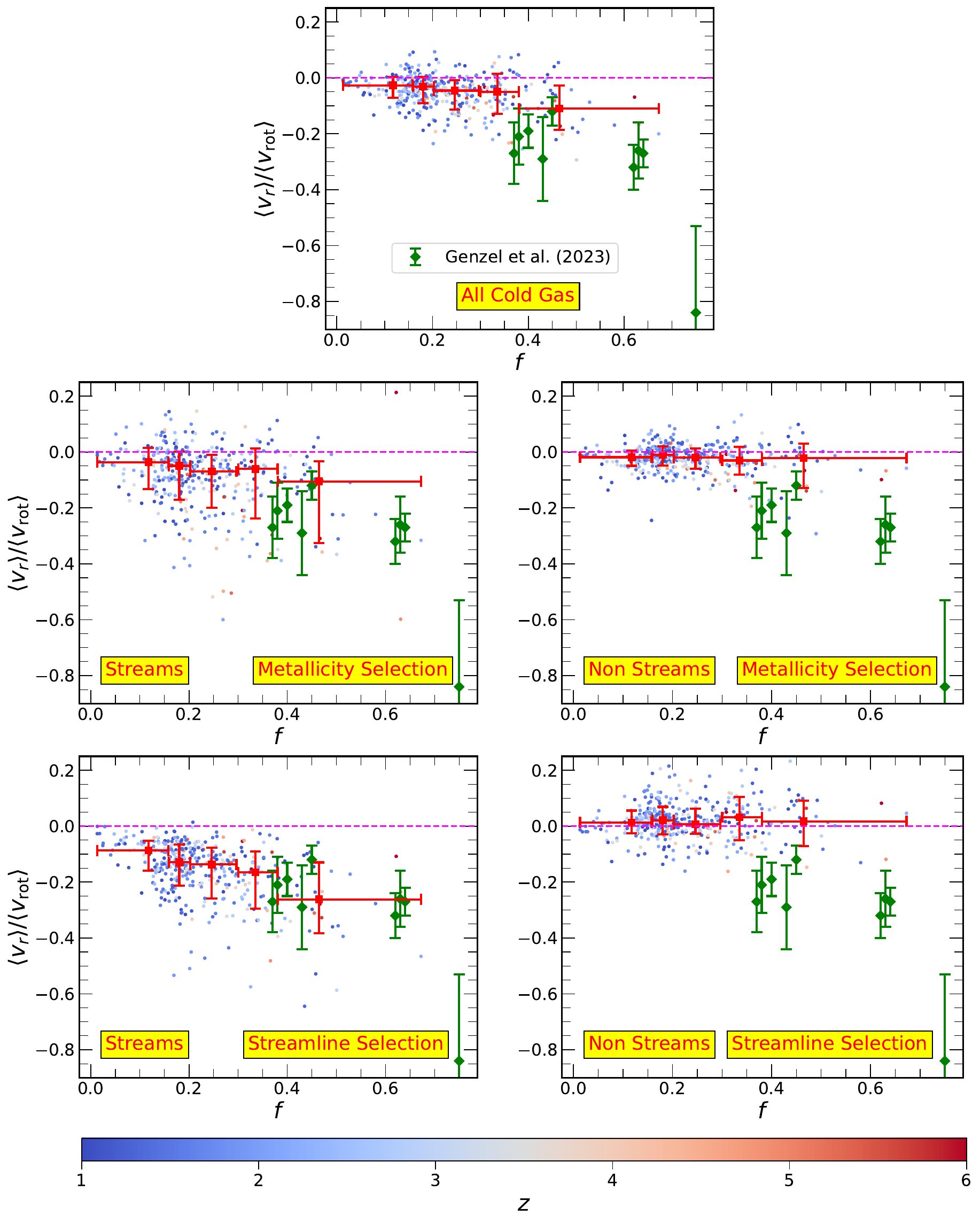}
    \caption{{\bf Comparison of Observations with the VELA Disk Galaxy Sample}: $\langle V_r \rangle / \langle V_{\rm rot} \rangle$ is plotted as a function of the cold gas fraction ($f$) for all cold gas (top panel), streams (middle and bottom left-hand panels), and non-streams (middle and bottom right-hand panels). Each circle represents a snapshot from the VELA disk galaxy sample and is color-coded by its corresponding redshift, $z$. Here, $\langle V_r \rangle$ denotes the stream-, non-stream-, or disk-averaged radial velocity, as relevant, and $\langle V_{\rm rot} \rangle$ is the disk-averaged rotational velocity. Red squares indicate the median values within bins of $f$, with vertical error bars showing the $16^{\rm th}$–$84^{\rm th}$ percentile ranges and horizontal bars reflecting the bin widths. The horizontal position of each square corresponds to the average $f$ within that bin. Results from the metallicity-based and streamline-based classifications are shown in the middle and bottom rows, respectively. For comparison, observational data from \citet{genzel23} are represented by green diamonds with vertical error bars. These observations are consistent with the stream-averaged radial velocities obtained from the simulations (middle and bottom left panels). In contrast, the disk-averaged (top row) and non-stream-averaged (middle and bottom right panels) radial velocities are generally of much smaller magnitude and show limited agreement with the observed values.}
    \label{fig:obs_sim_comp_vr_fg}
\end{figure*}

\begin{figure*}[h!!!]
    \centering
    \includegraphics[width=0.91\textwidth]{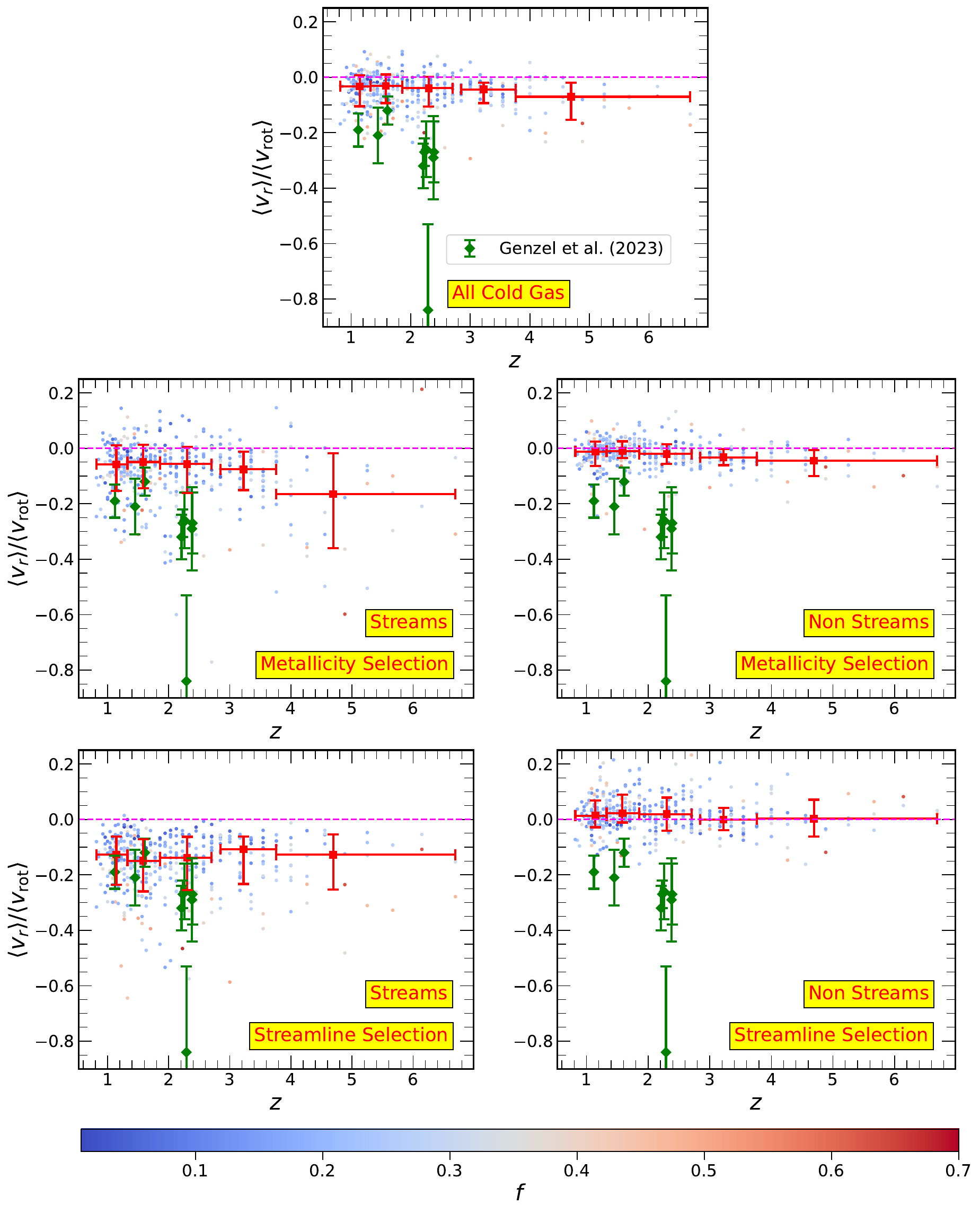}
    \caption{{\bf Comparison of Observations with the VELA Disk Galaxy Sample}: Same as in Figure~\ref{fig:obs_sim_comp_vr_fg}, except now $\langle V_r \rangle/\langle V_{\rm rot} \rangle$ is plotted against the redshift, $z$, and the circles corresponding to the simulated disks are color-coded by their gas fraction, $f$. The observations agree with the stream-averaged radial velocities found in the simulated disks (middle and bottom left-hand panels). The disk-averaged (top row) and non-stream-averaged (middle and bottom right-hand panels) radial velocities from the simulations typically have much smaller magnitudes and barely agree with the observations.}
    \label{fig:obs_sim_comp_vr_z}
\end{figure*}

\begin{figure*}[h!!!]
    \centering
    \includegraphics[width=0.9\textwidth]{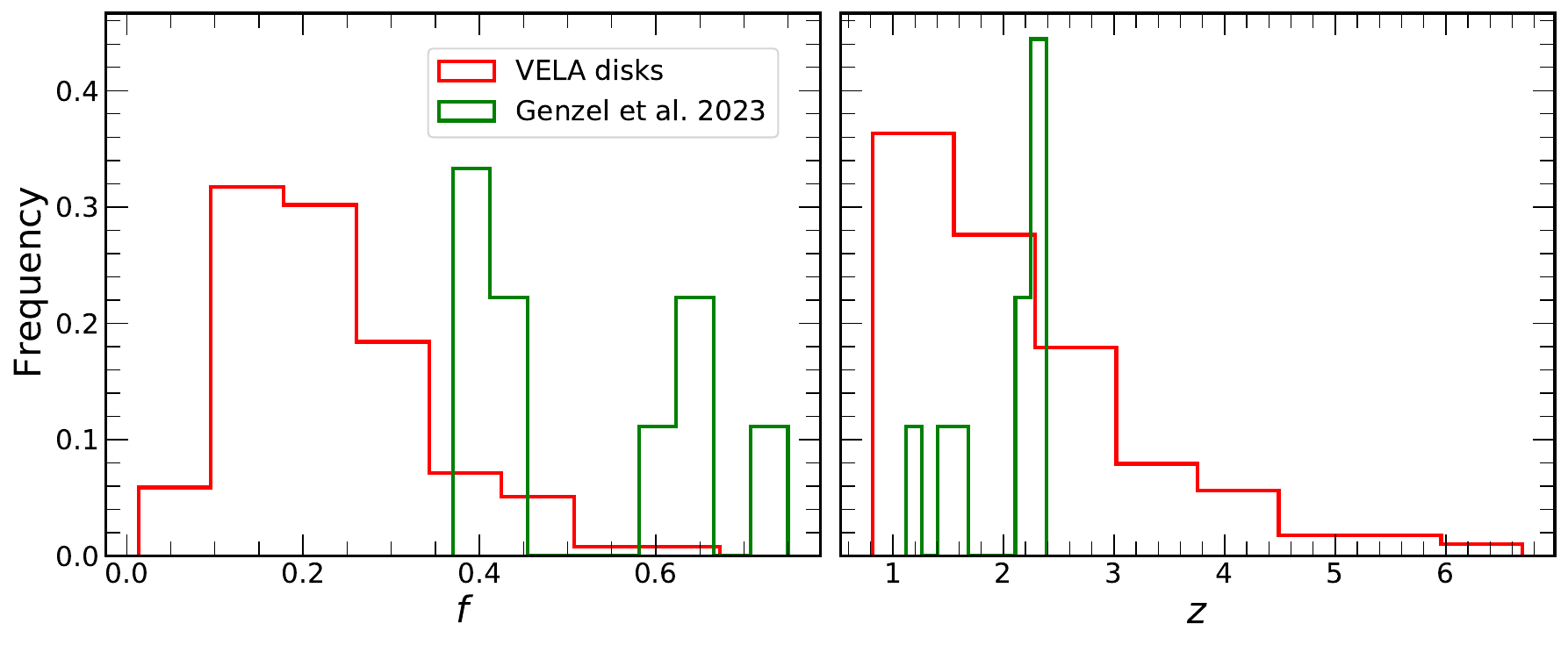}
    \caption{The left- and right-hand panels show the histograms for cold gas fraction, $f$, and redshift, $z$, for the VELA disks in red and the \citet{genzel23} galaxies in green. The VELA disks span a wider redshift range and have typically lower gas fractions than the \citet{genzel23} sample. While the lower gas fractions in VELA could potentially bias our conclusions, the simulated sample does include a few gas rich galaxies. But the disk-averaged radial velocities in even these gas-rich galaxies barely reproduce the observations, which are much better explained when assumed to be originating from recently accreted streams.}
    \label{fig:stats_genzel}
\end{figure*}

\section{Comparison with Observations}
\label{sec:compare_to_obs}

Using high-resolution H$\alpha$/CO imaging spectroscopy of nine rotating disks selected from the RC100 sample \citep[][]{genzel20,nestor23}, \citet{genzel23} find evidence for large non-circular motions. Assuming that all gas motions are in the plane of the disk, the non-circular motions are interpreted as radial. While distinguishing radial inflows from outflows can be difficult in observations because of projection effects, \citet{genzel23} argue that the inferred radial velocity signals from each of the nine disks are consistent with inflows \citep[see Section~2.4.1 and Table~3 of][]{genzel23}. The RC100 sample consists of massive star-forming galaxies near the star formation main sequence with redshifts around the peak of cosmic star formation activity ($0.6<z<2.5$). The nine selected galaxies are moderately large (half mass radius $> 4\ \rm kpc$), well-resolved (FWHM resolution $< 0.5 \arcsec$) with deep integrations ($> 10 \ \rm hr$), have low inclinations ($<70 \deg$), and are devoid of both strong AGN activity and perturbations from nearby massive companions. 

Using the forward modeling tool {\it DYSMAL} \citep[][]{genzel20, price21}, which includes the effects of beam smearing and instrumental resolution, \citet{genzel23} construct model velocity and velocity dispersion maps for purely rotating systems. After subtracting the model maps from the observed ones, second-order velocity residuals are obtained, which allow the study of radial motions. Independently, a kinemetry analysis \citep[][]{krajnovic06, shapiro08} of the observed velocity maps is also performed, providing model-independent evidence for large radial motions. The radial velocity signals obtained using the two different methods generally agree well with one another \citep[see Sections 2 and 3 of][]{genzel23}. 

\citet[][see Section 4.3.1]{genzel23} note that the observed radial velocities are in the same ballpark as the predictions from the different analytical models of radial transport based on disk instability (see \se{4} for a discussion on these models). However, given our current results from the VELA simulations that the radial transport of cold gas is dominated by the motion of recently accreted streams, the agreement between these model predictions and the observations may be a coincidence. 

Therefore, instead of using the disk-instability-driven transport models, here, we interpret the \citet{genzel23} observations in light of the VELA simulations. In Figs.~\ref{fig:obs_sim_comp_vr_fg} and \ref{fig:obs_sim_comp_vr_z}, $\langle V_r \rangle/\langle V_{\rm rot} \rangle$ is plotted against the cold gas fraction, $f$, and redshift, $z$, respectively, for all cold gas (top panel), streams (middle and bottom left-hand panels), and non-streams (middle and bottom right-hand panels), with each circle corresponding to a particular snapshot in the VELA disk galaxy sample, color-coded by $z$ in Fig.~\ref{fig:obs_sim_comp_vr_fg} and by $f$ in Fig.~\ref{fig:obs_sim_comp_vr_z}. Recall that $\langle V_r \rangle$ is the stream-, non-stream-, or disk-averaged radial velocity, as relevant, and $\langle V_{\rm rot} \rangle$ is the disk-averaged rotational velocity. The red squares and the vertical error bars denote the medians and $16^{\rm th}-84^{\rm th}$ percentiles, respectively, in $f$ and $z$ bins of varying widths, as indicated by the horizontal error bars, with the locations of the squares along the $f$- and $z$-axes being at the bin-averages. The results from the metallicity- and streamline-based selections are shown in the middle and bottom rows, respectively. The green diamonds with vertical error bars highlight the \citet{genzel23} observations, where $f$ corresponds to molecular gas fraction.

In Fig~\ref{fig:stats_genzel}, the left- and right-hand panels show the histograms for $f$ and $z$ for the \citet{genzel23} galaxies in green and the VELA disk galaxy sample in red. The \citet{genzel23} galaxies are located at redshifts between $1$ to $2.5$ and have large gas fractions ($f>0.35$). On the other hand, our disk galaxy sample spans a redshift range of $\sim \! 1$ to $\sim \! 6$ and has cold gas fractions varying from $\sim \! 0.01$ to $\sim \! 0.7$ (most of the sample has $f<0.35$). Nevertheless, where they overlap in $z$ and $f$, the \citet{genzel23} observations agree with the stream-averaged radial velocities from the simulations (middle and bottom left-hand panels of Figures~\ref{fig:obs_sim_comp_vr_fg} and \ref{fig:obs_sim_comp_vr_z}). Note that the stream-averaged radial velocities are largely similar in both selections, capable of attaining large negative values (up to $\sim \! 0.6-0.8$ times the disk-averaged rotational velocity). One exception is that the stream-averaged radial velocity in the streamline-based selection is always negative, whereas that in the metallicity-based selection can occasionally be positive (up to $\sim\! 0.2$ times the disk-averaged rotational velocity). The disk-averaged (top rows) and non-stream-averaged (middle and bottom right-hand panels) radial velocities from the simulations typically have much smaller magnitudes and barely agree with the observations.

We note that the VELA galaxies tend to have lower gas fractions than the observed galaxies. Given the generally increasing trend in the disk-averaged radial inflow levels with increasing gas fraction, this could qualitatively explain why the simulated median disk-averaged radial velocities are typically much lower than the observed values. However, the simulated sample does include a few gas-rich galaxies, with $f>0.35$ in $13 \%$ of the cases. The trend of disk-averaged radial velocity with gas fraction turns out to be rather weak, such that the disk-averaged radial velocities in these gas-rich galaxies barely reproduce the observed values, and it is much easier to explain the observations if assumed to be originating from freshly incoming streams. 

We thus conclude that the inflow velocities within the streams in the simulated disks may explain the radial velocities deduced from the observations if the observed signal is dominated by recently accreted streams and is characteristic of their (mass-weighted) average radial motions. A more robust comparison is yet to be performed, where observationally-realistic H$\alpha$ mocks of the simulated VELA disks are analyzed the same way as the observed galaxies using {\it DYSMAL}. Apart from accretion along cold streams, gas transport along a bar may be another way to achieve large radial motions \citep[][]{pastras25}. However, only one of the nine galaxies in the \citet{genzel23} sample shows strong evidence for the presence of a bar, and it is unclear whether bar-driven transport is also important for the rest of the sample.

\section{Conclusion}
\label{sec:7}

We studied the radial transport of cold gas in simulated star-forming disk galaxies at cosmic noon, $z \sim 4-1$, utilizing the VELA hydro-cosmological simulations, zooming in with $17.5 - 35 \pc$ maximum resolution on $34$ galaxies with halos of mass $10^{11}-10^{12} \Msun$ at $z \sim 2$. From all the VELA snapshots, rotation-supported disks were selected by requiring that the disk radius be at least three times the disk height and the average rotational velocity be greater than the radial velocity dispersion at all $r$, the cylindrical distance from the galactic center. With these selection criteria, the disk fraction among all VELA snapshots is $ \sim 39 \%$. Our main findings can be summarized as follows. 

\begin{itemize}

    \item[$\bullet$] The radial transport is dominated by inflows. \\
    
        \begin{itemize}
            \item The average radial velocity and radial mass flux as functions of $r$ are mostly negative and, on average, increase in magnitude with increasing $r$, except within $\sim \! 0.2\ R_\rmd$, where the trends are reversed. \\
            
            \item The disk-averaged radial velocities are negative in $82 \%$ of the galaxies and positive in the rest. \\

            \item The disk-averaged inflow is correlated with cold gas fraction and redshift. Both correlations are rather weak but more pronounced for the cold gas fraction. \\
        \end{itemize}
    
    \item[$\bullet$] For most of the VELA disks, the different analytical models of radial transport based on disk instability predict higher levels of disk-averaged inflow than that found in the simulations. An exception is the \citet{dekel20} model, which for $Q=1$ agrees reasonably well with the simulations. However, this model is not directly applicable to general disks, as it specifically refers to the migration of a thin ring. \\

    \item[$\bullet$] Using two different crude stream selection techniques, one based on metallicities and another on streamlines, the cold gas within the disk is classified into recently accreted streams versus off-stream material. On average, the streams are found to dominate the radial mass flux outside the inner disk ($r \gtrsim 0.3 R_\rmd$ for the streamline-based selection and r $\gtrsim 0.5 R_\rmd$ for the metallicity-based selection). \\

    \item[$\bullet$] The average radial velocities within the inflowing streams in the simulations are at the level of the inflow velocities indicated in the observed disks at cosmic noon by \citet{genzel23}. The simulated off-stream gas shows much lower average radial velocities. \\

\end{itemize}

We conclude that the inward radial transport of cold gas in extended, star-forming disk galaxies at cosmic noon is dominated by the external radial velocities of the incoming streams that penetrate the halo from the cosmic web and accrete onto the disk rather than being driven by torques from internal disk instabilities. Future studies with smoothed-particle hydrodynamics simulations or AMR simulations with tracer particles, where recently accreted gas along cold streams can be identified with better accuracy, should further solidify our novel result. We note that while the current ART code does not have the capability to add tracers, AMR simulations with tracers can be performed using other codes such as RAMSES \citep[][]{teyssier02}.

Our finding that the disk-averaged inward radial transport in the cosmological simulations is significantly weaker than the predictions of the analytical models for isolated disks poses an open question for future study. One option is that the dominant presence of anisotropic incoming streams from the cosmic web changes the overall conditions in the disks in a way that invalidates some of the assumptions made in the analytical models. In particular, it may affect the energy balance in the turbulent disk between streams, feedback, dissipation, and radial transport. Also, the premise that $Q$ is self-regulated to a value less than or equal to unity may not be valid for the simulated disks. Indeed, by analyzing a few of the VELA disks, \citet{inoue16} found that while collapsed clumps tend to have $Q<1$, proto-clump regions generally have $Q \gtrsim 2-3$. The high $Q$-values possibly represent excessive compressive modes of turbulence, likely induced by external perturbations, such as intense accretion along the incoming streams \citep[see also][]{mandelker24,ginzburg25}. However, there are also indications from simulations of isolated disks for low averaged radial inflows, with significant deviations from axial symmetry, qualitatively similar to the findings from our cosmological simulations (private communications with Andreas Burkert and Oscar Agertz). This may indicate weaknesses in the analytical models based on disk instability more profound than missing the incoming streams. It is also possible that contribution to gas turbulence-driving in such isolated disks by stellar feedback reduces the radial inflow rates compared to that predicted by the various disk-instability-based models ignoring feedback \citep[][]{krumholz18}. However, previous work by \citet{goldbaum16} using isolated disk simulations with gas fractions upto $20 \%$ show that including feedback only modestly lowers the levels of radial inflow compared to that in simulations without feedback \citep{goldbaum15}.

\begin{acknowledgements}
    We are grateful to the referee, Dylan Nelson, for many insightful comments, which led to significant improvements in the manuscript. We also thank Andreas Burkert, Joel Primack, and Oscar Agertz for stimulating discussions. This research has been supported by the Israel Science Foundation (ISF) grant 861/20, grants 2023723 and 2023730 from the US-Israel Binational Science Foundation and the US National Science Foundation (BSF-NSF), and the NSF grant PHY-2309135 to the Kavli Institute for Theoretical Physics (KITP). DDC acknowledges support from the Excellence Fellowship Program for International Postdoctoral Researchers by the Council for Higher Education and the Israel Academy of Sciences and Humanities. NM is supported by the ISF grant 3061/21 and the BSF-NSF grant 2020302. 
\end{acknowledgements}

\newpage

\bibliographystyle{aa}
\bibliography{ms} 

\begin{appendix}

\section{Impact of Different Feedback Models}
\label{sec:A1}

\begin{figure*}
    \centering
    \includegraphics[width=0.9\textwidth]{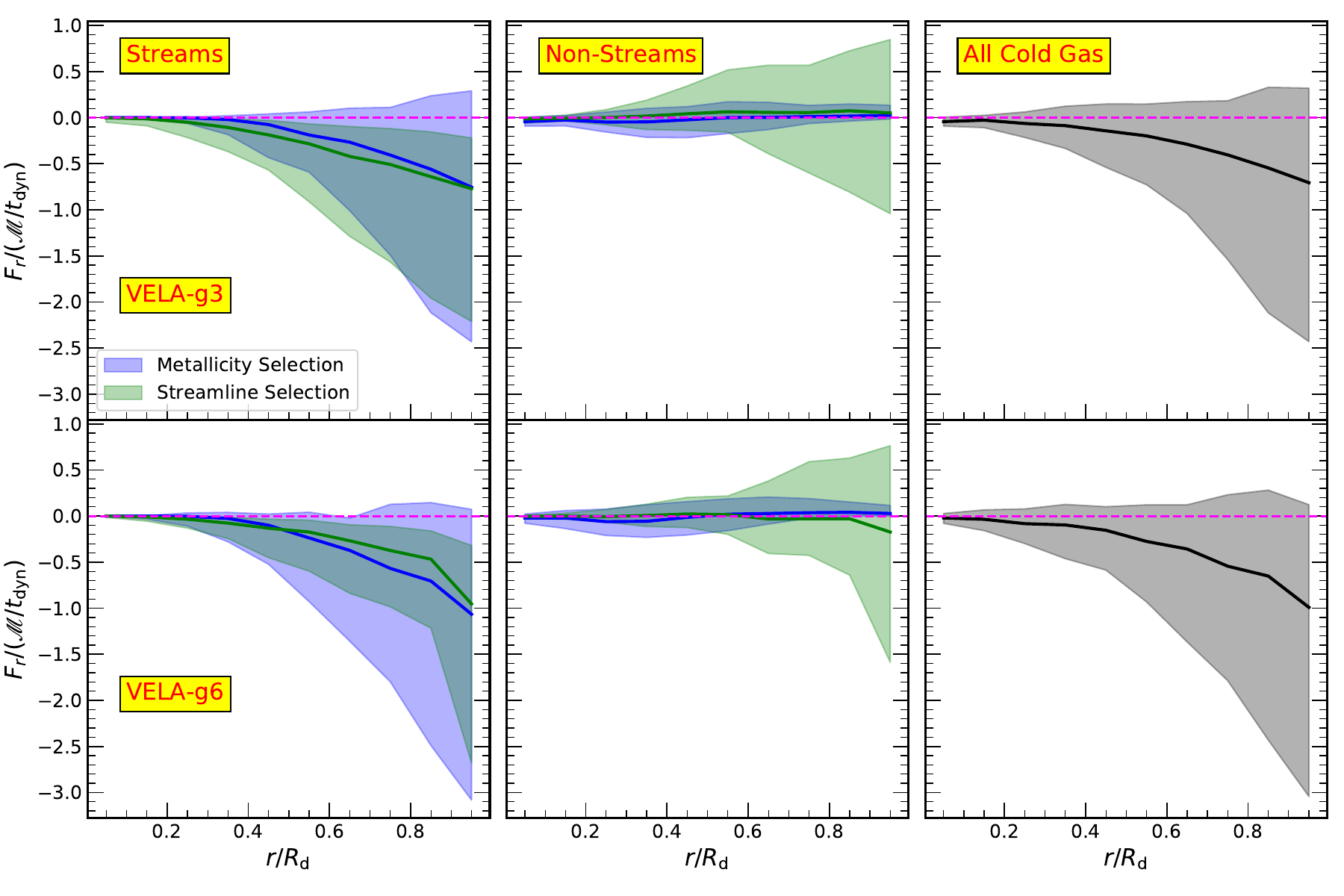}
    \caption{{\bf Cold Gas Radial Transport in VELA-g3 vs VELA-g6:} The top and bottom rows depict the radial mass flux, $F_r$ (in units of $\mathscr{M}/t_{\rm dyn}$), in VELA-g3 and VELA-g6 disks, respectively, with the left-hand, middle, and right-hand panels, respectively, focusing on streams, non-streams, and all cold gas. The solid lines indicate the medians and the envelopes represent the $16^{\rm th} - 84^{\rm th}$ percentile variations. Rotation-supported disks are identified from all snapshots using the procedure outlined in Section~\ref{sec:diskgs}, and the results from the metallicity- and streamline-based stream selections, as described in Sections~\ref{sec:met_sel} and \ref{sec:sl_sel}, are shown in blue and green, respectively. The median cold gas $F_r$ is slightly larger in magnitude in VELA-g6 disks, especially in the outer parts ($r> 0.5 R_\rmd$), likely due to somewhat larger gas fractions in VELA-g6 compared to VELA-g3 (this is because of stronger stellar feedback in VELA-g6, see Fig.~\ref{fig:gas_frac}). However, $F_r$ is dominated by stream inflows in both VELA-g3 and VELA-g6 disks, especially in the outer parts, irrespective of the strength of stellar feedback or the details of feedback physics.}
    \label{fig:difffeedback}
\end{figure*}

\begin{figure}
    \centering
    \includegraphics[width=0.45\textwidth]{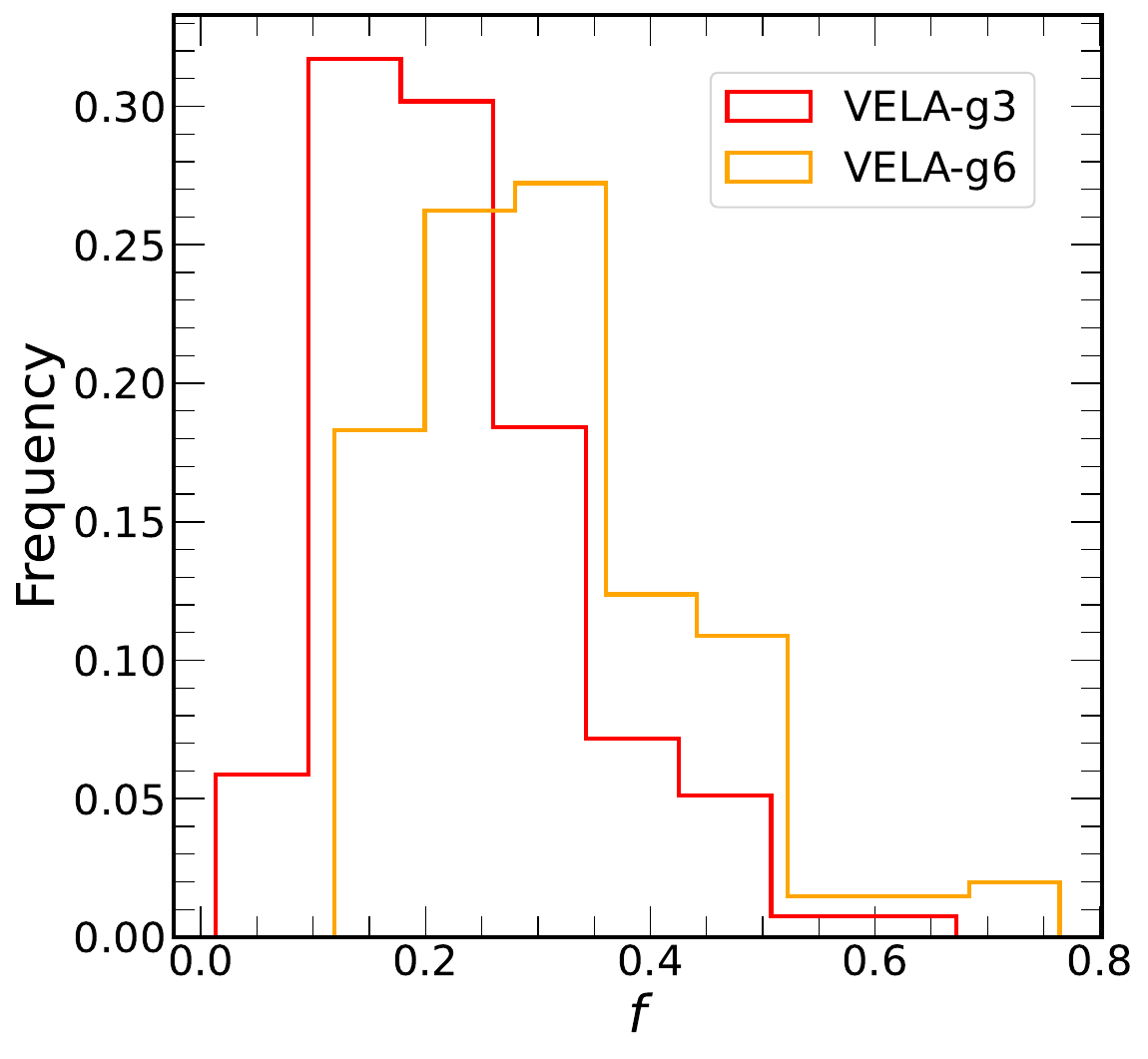}
    \caption{{\bf Cold Gas Fractions in VELA-g3 vs VELA-g6:} Histograms for the cold gas fractions in VELA-g3 and VELA-g6 disks are shown in red and orange, respectively. On average, gas fractions in VELA-g6 disks tend to be somewhat higher, with a median value of $0.3$ as opposed to $0.2$ in VELA-g3 disks.}
    \label{fig:gas_frac}
\end{figure}

The results discussed in the main text are derived from analyzing the VELA simulations introduced in \citet{ceverino14} and \citet{zolotov15}, hereafter referred to as VELA-g3. A newer generation of the VELA simulations was presented in \citet{ceverino23}, which, referred to as VELA-g6, uses the same set of initial conditions as in VELA-g3 but differs in its modeling of stellar feedback. In addition to thermal feedback from supernovae and feedback from radiation pressure (same as that in VELA-g3), VELA-g6 includes moderate IR trapping and kinetic feedback from expanding supernovae shells and stellar winds, resulting in stronger stellar feedback.

Here, we ascertain the impact of the different feedback models used in VELA-g3 and VELA-g6, if any, on the properties of the radial transport of cold gas in the simulated disks. Rotation-supported disks are identified from the VELA-g6 snapshots using the same procedure as that adopted for VELA-g3 (see Section~\ref{sec:diskgs}). The cold gas in these disks is then classified into streams and non-streams following the same metallicity and streamline-based selection techniques and thresholds used for VELA-g3, as discussed in Sections~\ref{sec:met_sel} and \ref{sec:sl_sel}, respectively.

In Fig.~\ref{fig:difffeedback}, the left-hand, middle, and right-hand panels show the radial mass flux, $F_r$ (in units of of $\mathscr{M}/t_{\rm dyn}$), for streams, non-streams, and all cold gas, respectively, for the VELA-g3 (top row) and VELA-g6 (bottom row) disk galaxy samples,  with the solid lines indicating the medians and the envelopes representing the $16^{\rm th} - 84^{\rm th}$ percentile variations. The results from the metallicity- and streamline-based selections are shown in blue and green, respectively.

Upon comparing the results from the two generations, the median cold gas $F_r$ in VELA-g6 disks is found to be slightly larger in magnitude than that in VELA-g3 disks, especially in the outer parts ($r> 0.5\ R_\rmd$). This is likely because, on average, gas fractions in VELA-g6 are somewhat higher than that in VELA-g3 (a weak correlation between the strength of radial transport and cold gas fraction in disks is established in Section~\ref{sec:stats}), an effect of stronger stellar feedback \citep[see][]{ceverino23}. For reference, the histograms for the cold gas fractions, $f$, in VELA-g3 and VELA-g6 disks are shown in Fig.~\ref{fig:gas_frac} in red and orange, respectively. While the median and $16^{\rm th} - 84^{\rm th}$ percentile variation in $f$ for VELA-g3 disks is $0.20^{+0.13}_{-0.06}$, that for VELA-g6 disks is $0.30^{+0.14}_{-0.10}$. However, despite the differences discussed above, Fig.~\ref{fig:difffeedback} shows that the radial transport of cold gas in both VELA-g3 and VELA-g6 disks, especially in the outer parts, is dominated by stream inflows or the motion of recently accreted gas.

Thus, our key result, namely, the dominance of stream inflows in the radial transport of cold gas in simulated, cosmic noon disks, holds irrespective of the strength of stellar feedback or the details of the feedback physics.

\section{Impact of Different Stream Selections}
\label{sec:A2}

\begin{figure*}
    \centering
    \includegraphics[width=0.9\textwidth]{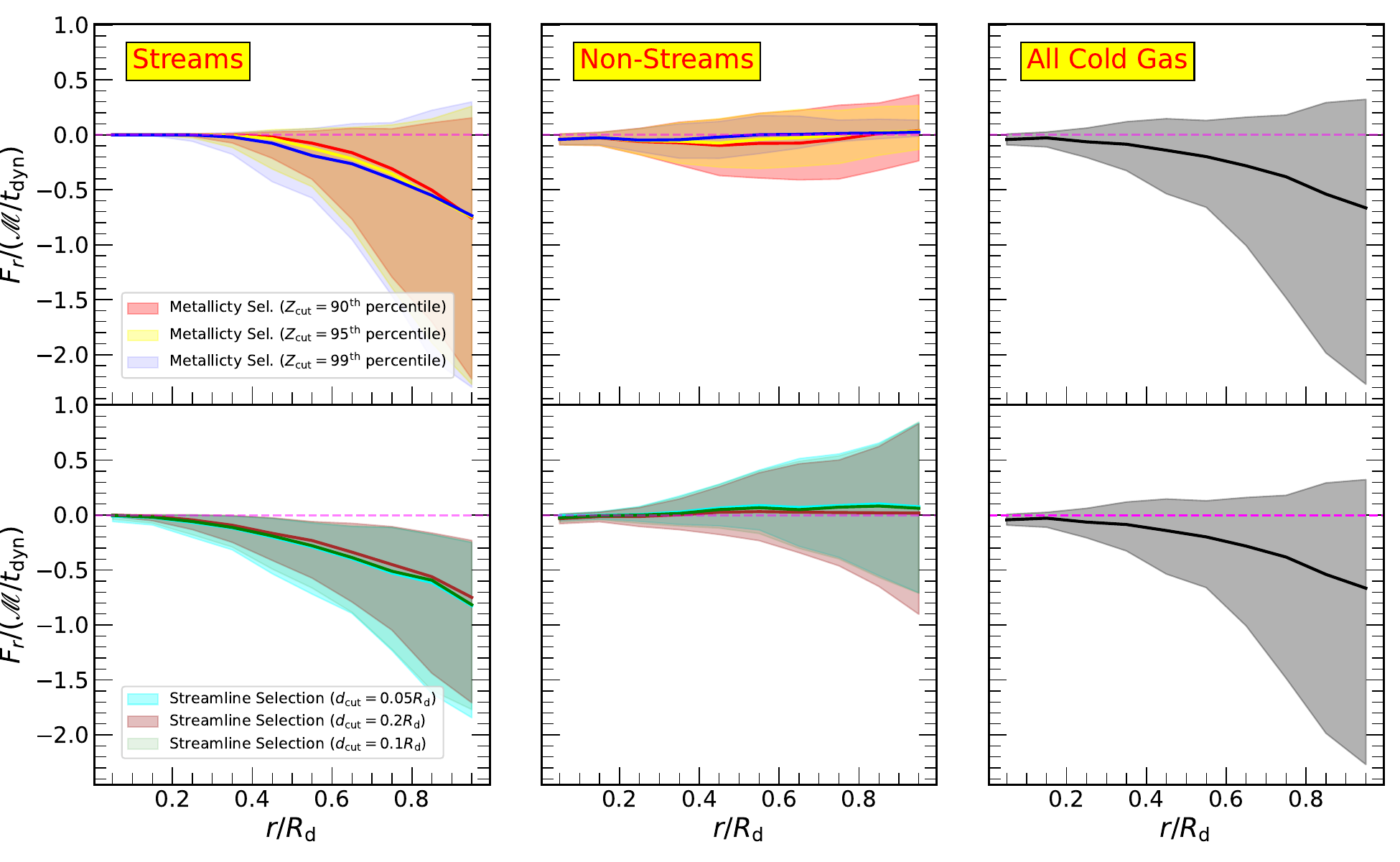}
    \caption{{\bf Cold Gas Radial Transport in VELA-g3 with Different Stream Selection Thresholds:} The top and bottom rows show the radial mass flux, $F_r$ (in units of $\mathscr{M}/t_{\rm dyn}$) in the metallicity- and streamline-based selections, respectively, applied to VELA-g3 disks, with the left-hand, middle, and right-hand panels, respectively, focusing on streams, non-streams, and all cold gas. The solid lines indicate the medians and the envelopes represent the $16^{\rm th} - 84^{\rm th}$ percentile variations. Results are shown for $Z_{\rm cut}=p^{\rm th}$ percentile, where $p=99$ (blue), $95$ (yellow), and $90$ (red), and $d_{\rm cut}=0.05$ (cyan), $0.1$ (green), and $0.2 R_\rmd$ (brown). Here,  $Z_{\rm cut}$ is the threshold metallicity from the mass-weighted metallicity distribution between $r=R_\rmd$ to $2 R_\rmd$ and $z=-H_\rmd$ to $H_\rmd$ used to distinguish between streams and non-streams in the metallicity-based selection (see Section~\ref{sec:met_sel}). Similarly, $d_{\rm cut}$ is the minimum distance by which gas cells must travel inwards over the average disk dynamical time to be classified as streams in the streamline-based selection (see Section~\ref{sec:sl_sel}). While the median stream signal is slightly larger in magnitude for larger values of $Z_{\rm cut}$ and smaller values of $d_{\rm cut}$, the dominance of incoming streams in the cold gas radial mass flux is insensitive to the exact values adopted for the thresholds in the different stream selection procedures.}
    \label{fig:diffselect}
\end{figure*}

As the VELA simulations are Eulerian in nature, identifying incoming streams of cold gas is a non-trivial task. In Sections~\ref{sec:met_sel} and \ref{sec:sl_sel}, we, respectively, describe the crude procedures adopted in this paper based on gas metallicities and streamlines for selecting cold gas cells that may belong to recently accreted streams. Here, we investigate whether variations in the default thresholds adopted in these methods to distinguish between streams and non-streams significantly affect our main result.

In the metallicity-based selection, the metallicity at the $p^{\rm th}$ percentile ($Z_{\rm cut}$) of the mass-weighted metallicity distribution between $r=R_\rmd$ to $2 R_\rmd$ and $z=-H_\rmd$ to $H_\rmd$ is used to distinguish between streams and non-streams (see Section~\ref{sec:met_sel}). Here, we study the dependence of stream selection on $Z_{\rm cut}$ by adopting different values of $p$, namely, $90$, $95$, and $99$, with $p=99$ being the default used in the main text. Similarly, in the streamline-based selection, gas cells that move inwards by distances greater than $d_{\rm cut}$ over the average disk dynamical time are classified as streams and the rest as non-streams (see Section~\ref{sec:sl_sel}). We also study the sensitivity of stream selection on $d_{\rm cut}$ by allowing it to vary within a factor of $2$ ($0.05$ or $0.2 R_\rmd$) of the default value of $0.1 R_\rmd$ used in the main text.

In Figure~\ref{fig:diffselect}, the left-hand, middle, and right-hand panels show the radial mass flux, $F_r$ (in units of $\mathscr{M}/t_{\rm dyn}$), for streams, non-streams, and all cold gas, respectively, for the metallicity- (top row) and streamline-based (bottom row) selections with different thresholds, as labeled, and applied to VELA-g3 disks. The solid lines indicate the medians, and the envelopes represent the $16^{\rm th} - 84^{\rm th}$ percentile variations. As expected, the median stream signal is slightly larger in magnitude for larger values of $Z_{\rm cut}$ and smaller values of $d_{\rm cut}$, which tend to select more mass in streams. However, irrespective of the values adopted for $Z_{\rm cut}$ and $d_{\rm cut}$, the cold gas radial mass flux is dominated by the stream flux, especially in the outer disks.

Thus, our main result pertaining to the dominance of stream inflows in the radial transport of cold gas in simulated, cosmic noon disks holds independently of the exact values adopted for the thresholds in the different stream selection procedures.

\section{Comparison with Transport Models at Different Radii}
\label{sec:A3}

\begin{figure*}
    \includegraphics[width=0.95\textwidth]{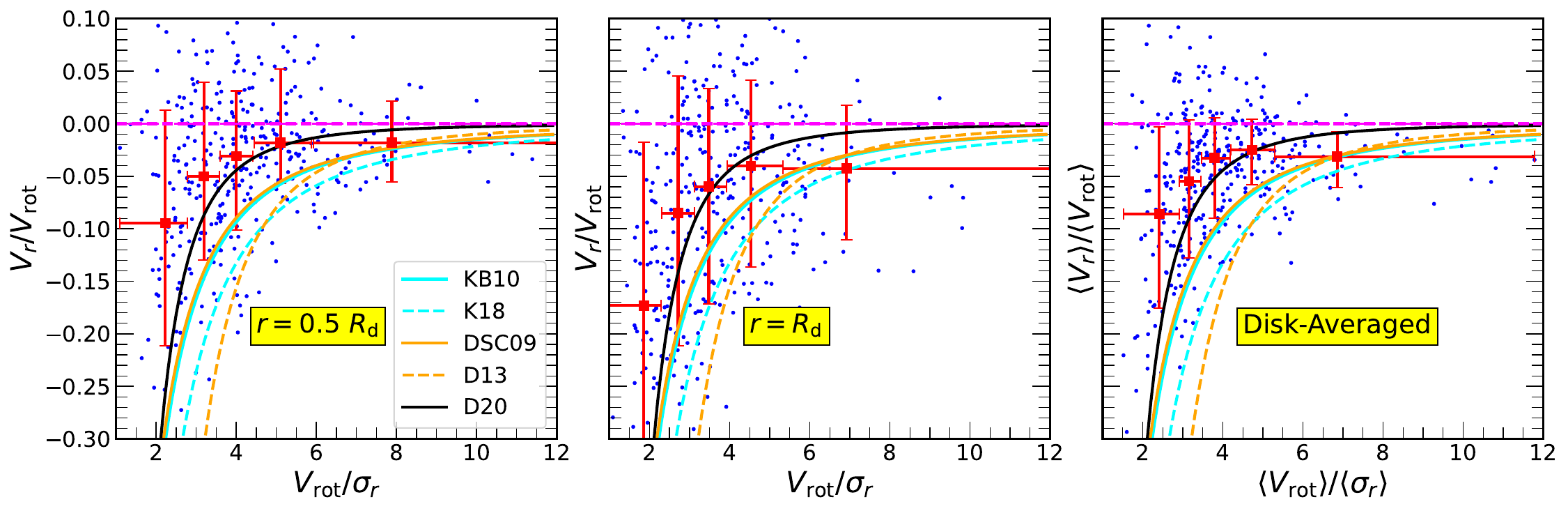}
    \caption{{\bf Comparison with Disk-Instability-Based Transport Models at Different Radii:} For each of the VELA disks, the left-hand and middle panels plot $V_r/V_{\rm rot}$ against $V_{\rm rot}/\sigma_r$ at $r=0.5\ R_{\rm d}$ and $r=R_{\rm d}$, respectively, with the blue, round points. The corresponding disk-averaged quantities are shown in the right-hand panel. Red squares indicate the median values within bins of $x$ (where $x=V_{\rm rot}/\sigma_r$ or $\langle V_{\rm rot} \rangle/\langle \sigma_r \rangle$, as applicable), with the vertical error bars highlighting the $16^{\rm th}-84^{\rm th}$ percentile ranges and the horizontal error bars showing the bin widths. The disk-instability-based model predictions (see Section~\ref{sec:4}) are shown with the different curves, as labeled, assuming $Q=1$. For most of the VELA disks, these models predict higher levels of inflow than that seen in the simulations, not only in a disk-averaged sense but also locally at different $r$.
    }
    \label{fig:radial_dep}
\end{figure*}

In the main text, we demonstrate that the predictions from the disk-instability-based radial transport models are a poor match to the disk-averaged radial inflow velocities seen in most of the VELA disks. Here, we investigate whether the same also holds when comparing the average inflow velocities at different radii.

The first two panels of Figure~\ref{fig:radial_dep} plot $V_r/V_{\rm rot}$ against $V_{\rm rot}/\sigma_r$ for each of the VELA disks with the blue, round points, the left-hand and middle panels corresponding to $r=0.5\ R_{\rm d}$ and $r=R_{\rm d}$, respectively. The corresponding disk-averaged quantities are plotted against each other in the right-hand panel. In each panel, the red squares and the vertical error bars denote the medians and the $16^{\rm th}-84^{\rm th}$ percentiles, respectively, in $x$ bins of varying widths (where $x=V_{\rm rot}/\sigma_r$ or $\langle V_{\rm rot} \rangle/\langle \sigma_r \rangle$, as applicable), with the locations of the squares along the $x$-axes being at the bin-averages. The predictions for the radial inflow velocities from the various disk-instability-based transport models are overlaid with the different curves, as labeled, assuming $Q=1$. Most of the VELA disks show low average inflow velocities, not-only in a disk-averaged sense, but also both at $r=0.5\ R_{\rm d}$ and $r=R_{\rm d}$, compared to that predicted by the different transport models. We have verified this is true for other radial bins as well. Assuming $Q=0.68$ in the models would result in an even poorer match, as the model inflows increase as $Q$ decreases (see Section~\ref{sec:4}).

\section{Correlation of Radial Transport with Cold Gas Mass}
\label{sec:A4}

\begin{figure}
    \includegraphics[width=0.5\textwidth]{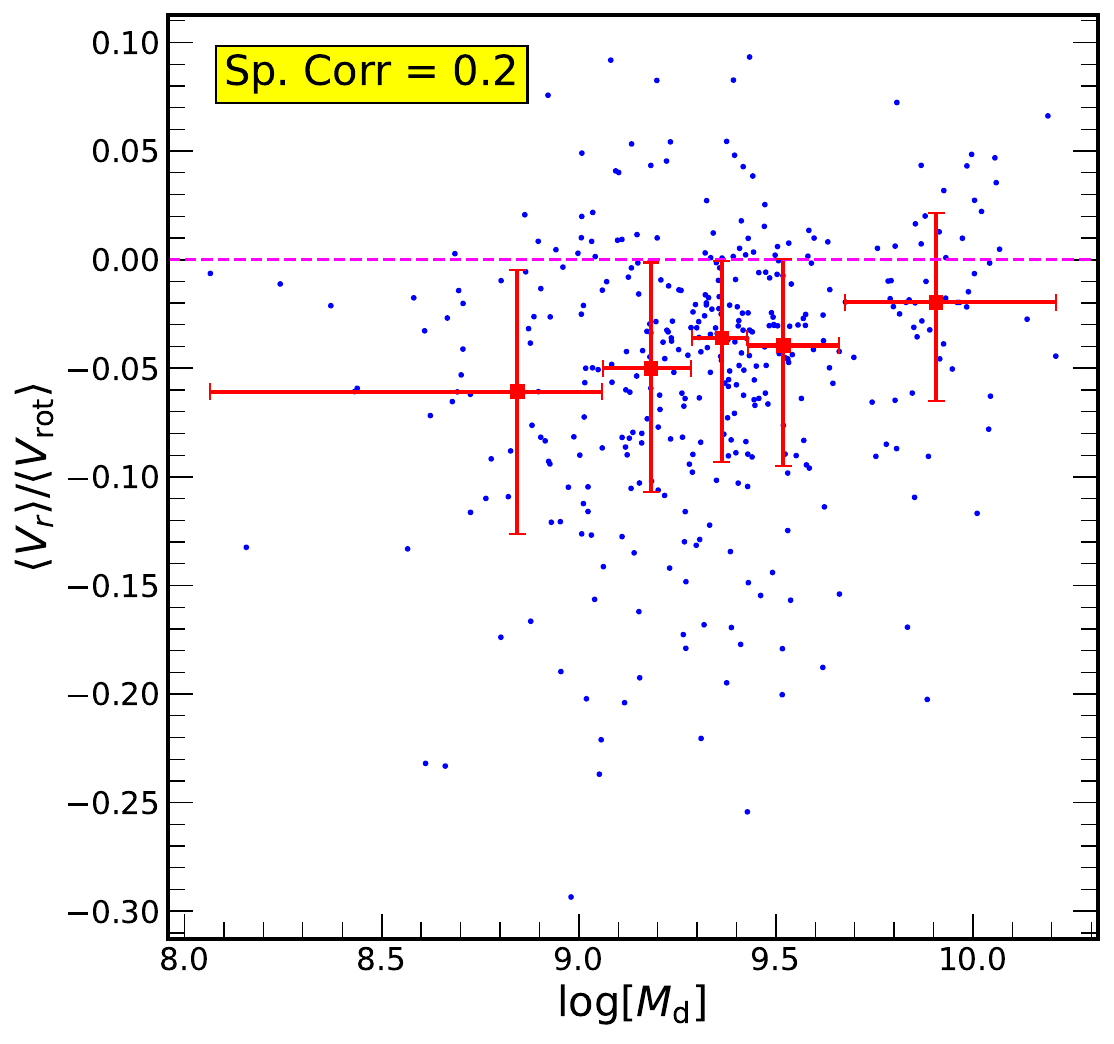}
    \caption{{\bf Trend in Disk-Averaged Inflow with Cold Gas Mass}: The ratio of the disk-averaged radial to rotational velocity, $\langle V_r \rangle/\langle V_{\rm rot} \rangle$, is plotted against the total cold gas mass in the disk, $M_{\rm d}$ (in logarithmic units), with the blue, round points for the VELA disk galaxy sample. Red squares indicate the median values within bins of log[$M_{\rm d}$], with the vertical error bars highlighting the $16^{\rm th}-84^{\rm th}$ percentile ranges and the horizontal error bars showing the bin widths. On average, with increase in $M_{\rm d}$, the magnitude of $\langle V_r \rangle/\langle V_{\rm rot} \rangle$ slightly decreases. The correlation, however, is quite weak, with a Spearman rank correlation coefficient of $0.2$.}
    \label{fig:mass_corr}
\end{figure}

Here, we examine the correlation, if any, of the radial transport of cold gas through the disk with the total cold gas mass.

In Figure~\ref{fig:mass_corr}, $\langle V_r \rangle/\langle V_{\rm rot} \rangle$ is plotted against $M_{\rm d}$ (in logarithmic units), the total cold gas mass in the disk, for each of the VELA disks with the blue, round points. The red squares and the vertical error bars denote the medians and the $16^{\rm th}-84^{\rm th}$ percentiles, respectively, in bins of log[$M_{\rm d}$], with the locations of the squares along the log[$M_{\rm d}$] axis being at the bin-averages. We find a very weak trend with log[$M_{\rm d}$], bearing a Spearman rank correlation coefficient of $0.2$. On average, as $M_{\rm d}$ increases, the magnitude of $\langle V_r \rangle/\langle V_{\rm rot} \rangle$ slightly decreases.

\section{Radial Transport of Warm/Hot Gas}
\label{sec:A5}

\begin{figure}
    \centering
    \includegraphics[width=0.5\textwidth]{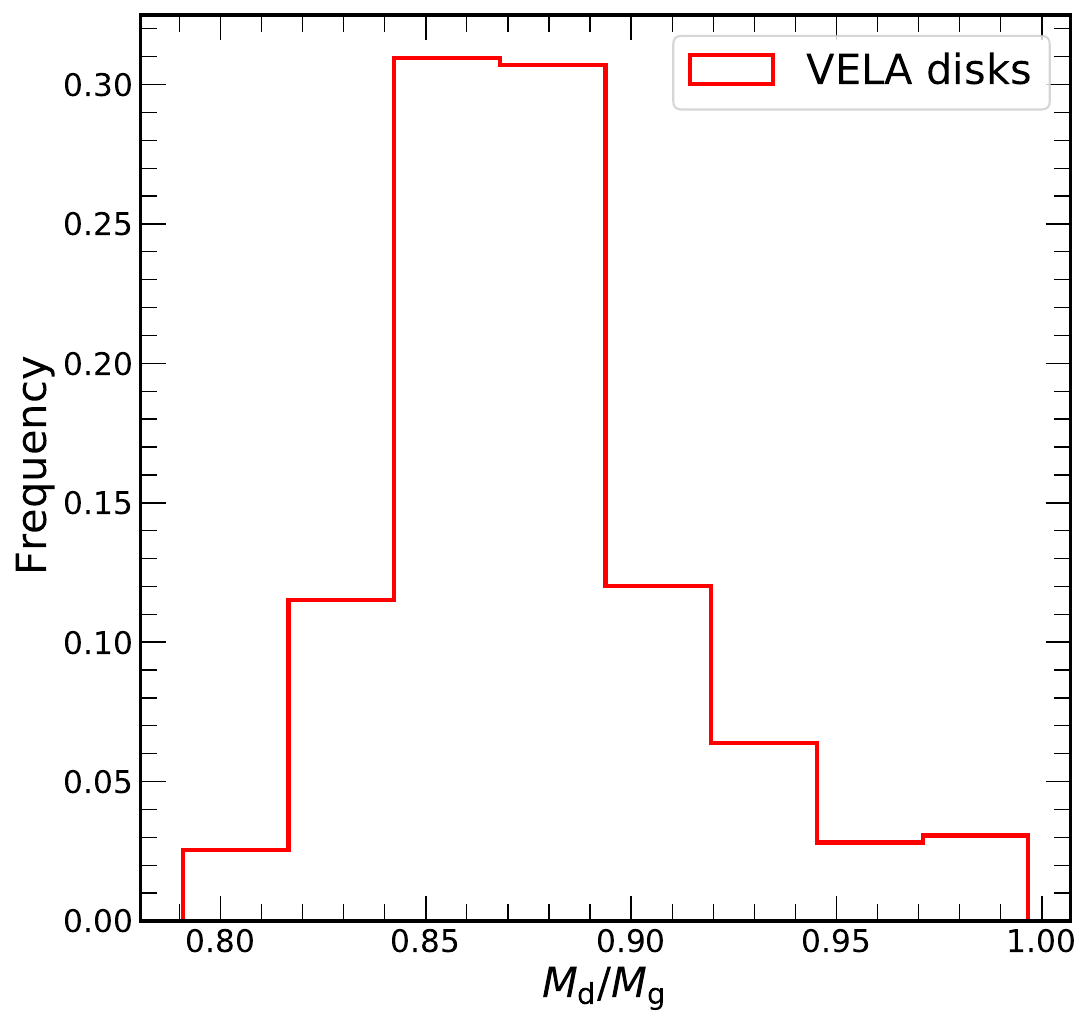}
    \caption{{\bf Mass Fraction of Cold Gas to Total Disk Gas}: Histogram for the ratio of the mass in the cold phase ($T<1.5 \times 10^{4} {\rm K}$), $M_{\rm d}$, to the total gas mass in the disk, $M_{\rm g}$. Atleast $80 \%$ and typically, around $85$ to $90 \%$ of the disk gas mass is in the cold phase.}
    \label{fig:cold_to_all}
\end{figure}

\begin{figure*}
    \centering
    \includegraphics[width=0.9\textwidth]{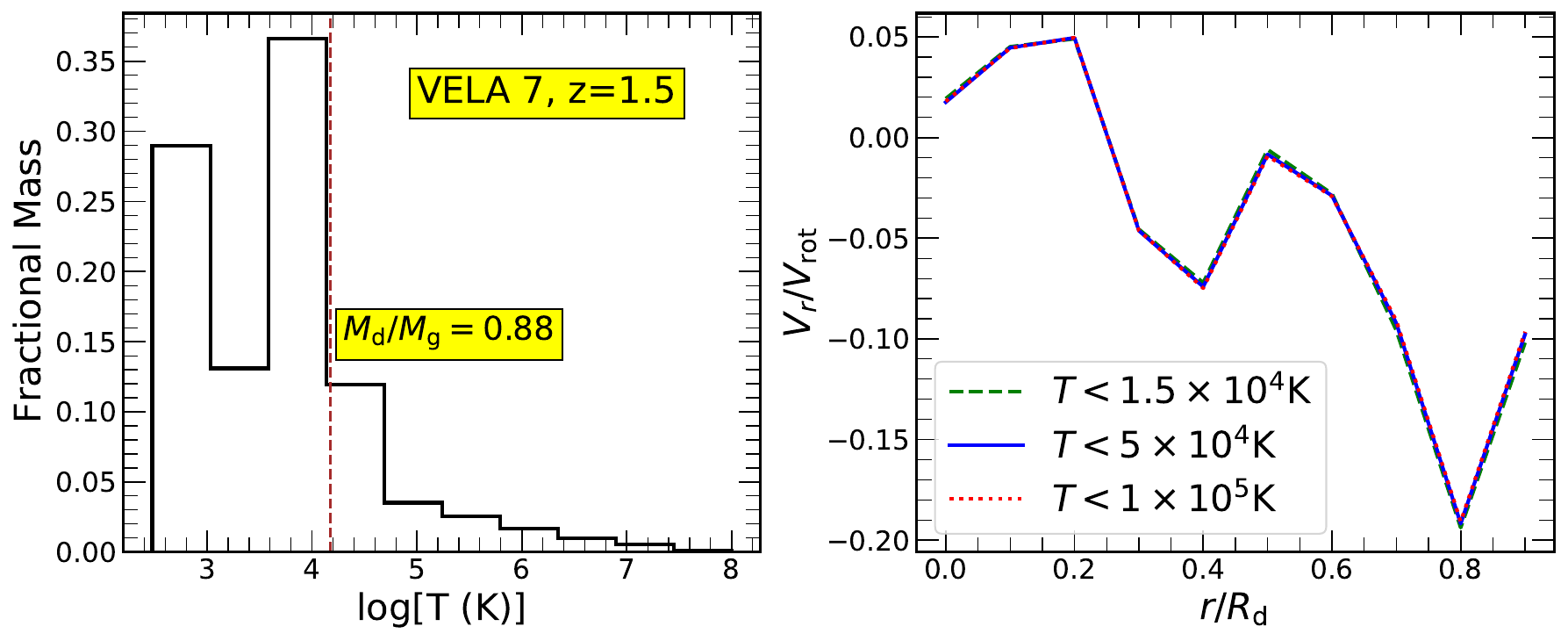}
    \caption{{\bf Radial Transport through the Disk for Different Gas Temperature Thresholds}: Left-hand panel shows the mass-weighted histogram for the disk gas temperatures for VELA 7 at $z=1.5$. The brown, dashed line indicates $T=1.5 \times 10^{4} {\rm K}$, the threshold used to define the cold phase, and the cumulative mass in the cold phase is around $88 \%$ of the total gas mass in the disk. Right-hand panel plots $V_{r}/V_{\rm rot}$, as a function of $r$ (in units of $R_{\rm d}$) for three different gas temperature cuts, as labeled. There is no significant additional contribution to average radial velocities through the disk from gas above $T=1.5 \times 10^{4} {\rm K}$.}
    \label{fig:temp_dis}
\end{figure*}

Here, we investigate if there is any significant, additional contribution to the radial transport through VELA disks from gas above $1.5 \times 10^{4} {\rm K}$.

In Figure~\ref{fig:cold_to_all}, we show the histogram for the ratio of the cold gas (defined as gas below $1.5 \times 10^{4} {\rm K}$) mass, $M_{\rm d}$, to the total gas mass in the disk, $M_{\rm g}$, for the VELA disk galaxy sample. At least $80 \%$ and typically around $85$-$90 \%$ of the gas mass in the disks is in the cold phase. Hence, we do not expect any additional major contribution to the radial transport through the disk from gas above $1.5 \times 10^{4} {\rm K}$.

In Figure~\ref{fig:temp_dis}, we demonstrate this explicitly for VELA-7 at $z=1.5$. The left-hand panel shows the mass-weighted histogram for disk gas temperatures. The cumulative mass in the cold phase is around 88 $\%$ of the total gas mass in the disk. The right-hand panel shows the ratio of the average radial-to-rotational velocity in cylindrical shells, $V_{r}/V_{\rm rot}$, as a function of $r$ (in units of $R_{\rm d}$), adopting three different gas temperature cuts, as labeled. We find no significant additional contribution to the radial transport through the disk from gas above $1.5 \times 10^{4} {\rm K}$.

\end{appendix}

\end{document}